\documentclass[
reprint,
superscriptaddress,
nofootinbib,
amsmath,amssymb,
pra,
]{revtex4-2}
\usepackage{graphicx}
\usepackage{dcolumn}
\usepackage{hyperref}
\usepackage{subfig}
\usepackage{ragged2e}
\usepackage{booktabs}
\usepackage{bm}

\begin{document}

\preprint{APS/123-QED}

\title{Hybrid quantum algorithms for flow problems}

\author{Sachin S. Bharadwaj}
\email{sachin.bharadwaj@nyu.edu}
 \affiliation{Department of Mechanical and Aerospace Engineering, New York University, New York, NY 11201 USA}
 
\author{Katepalli R. Sreenivasan}
\email{katepalli.sreenivasan@nyu.edu}
 \affiliation{Department of Mechanical and Aerospace Engineering, New York University, New York, NY 11201 USA}
\affiliation{Courant Institute of Mathematical Sciences, New York University, New York, NY 10012}
 \affiliation{Department of Physics, New York University, New York, NY 10012}
 \affiliation{Center for Space Science, New York University Abu Dhabi, Abu Dhabi 129188, United Arab Emirates}

\date{\today}

\begin{abstract}
\noindent For quantum computing (QC) to emerge as a practically indispensable computational tool, there is a need for quantum protocols with an end-to-end practical applications---in this instance, fluid dynamics. We debut here a high performance quantum simulator which we term \textit{QFlowS} (Quantum Flow Simulator), designed for fluid flow simulations using QC. Solving nonlinear flows by QC generally proceeds by solving an equivalent infinite dimensional linear system as a result of linear embedding. Thus, we first choose to simulate two well known flows using QFlowS and demonstrate a previously unseen, full gate-level implementation of a hybrid and high precision Quantum Linear Systems Algorithms (QLSA) for simulating such flows at low Reynolds numbers. The utility of this simulator is demonstrated by extracting error estimates and power law scaling that relates $T_{0}$ (a parameter crucial to Hamiltonian simulations) to the condition number $\kappa$ of the simulation matrix, and allows the prediction of an optimal scaling parameter for accurate eigenvalue estimation. Further, we include two speedup preserving algorithms for (a) the functional form or sparse quantum state preparation, and (b) \textit{in-situ} quantum post-processing tool for computing nonlinear functions of the velocity field. We choose the viscous dissipation rate as an example, for which the end-to-end complexity is shown to be $\mathcal{O}(\textrm{polylog} (N/\epsilon)\kappa/\epsilon_{QPP})$, where $N$ is the size of the linear system of equations, $\epsilon$ is the the solution error and $\epsilon_{QPP}$ is the error in post processing. This work suggests a path towards quantum simulation of fluid flows, and highlights the special considerations needed at the gate level implementation of QC.
\end{abstract}

\maketitle
\section{Introduction}
\label{sec:Intro}
Computer simulations of nonlinear physical systems---such as turbulent flows, glassy systems, climate physics, molecular dynamics and protein folding---are formidably hard to perform on even the most powerful supercomputers of today or of foreseeable future.
In particular, the state-of-the-art Direct Numerical Simulations (DNS) of turbulent flows \cite{iyer2020classical,yeung2020advancing,yeung2022turbulence} governed by the Navier-Stokes equations, or of turbulent reacting flow problems and combustion \cite{rood2021enabling}, both of which involve massive simulations with high grid resolutions, not only reveal fine details of the flow physics \cite{iyer2021area,buaria2022scaling}, but also constantly contend with the limits of supercomputers on which the codes run \cite{bell2017look}. However, simulation sizes required to settle fundamental asymptotic theories, or simulate turbulent systems such as the Sun or cyclones, or to simulate flows around complex geometries of practical interest, would require computing power that is several orders of magnitude higher than is currently available. Reaching such computational {targets} calls for a paradigm shift in the computing technology. 

One such potential candidate is Quantum Computing (QC)\cite{nielsen2002quantum},
which has striven to establish its advantage over classical counterparts by promising polynomial or exponential speedups \cite{jordan}. Even though QC has been around for the last two decades, the subject is still nascent. In this nascent era, which has been called the Noisy Intermediate Scale Quantum (NISQ) era, QC's applications already extend \cite{awschalom2022roadmap,alexeev2021quantum} across finance, chemistry, biology, communication and cryptography, but not as much in areas that are governed by nonlinear partial differential equations, such as fluid dynamics.

This work attempts to pave the way for utilizing QC in Computational Fluid Dynamics (CFD) research, which we have termed \cite{bharadwaj2020quantum} Quantum Computation of Fluid Dynamics (QCFD). An initial comprehensive survey of various possible directions of QCFD was made in \cite{bharadwaj2020quantum}. Realistic CFD simulations with quantum advantages require one to quantumly solve general nonlinear PDEs such as the Navier-Stokes equations. However, it is worth noting that the fundamental linearity of quantum mechanics itself blockades encoding of nonlinear terms, thus forcing a linearization of some kind \cite{liu2021efficient,lin2022koopman,giannakis2022embedding}, which typically results in an \textit{infinite} dimensional linear system. In such cases the inaccessibility to the required large number of qubits (and thus exponentially large vector spaces) leads to inevitable truncation errors, limiting the focus to weakly nonlinear problems \cite{liu2021efficient}. Therefore the ability to solve high dimensional linear systems in an end-to-end manner\footnote{ By this we mean an algorithm that efficiently prepares a quantum state, processes it and outputs a result by measurement while retaining all or some net quantum advantage. } while capturing the flow physics is crucial to simulating nonlinear flow problems. Our goal here is to present various steps involved in the process of solving simple and idealized problems, including providing estimates of scaling and errors involved.

To achieve this goal, we present here a high performance quantum simulator which we call \textit{QFlowS}(Quantum Flow Simulator), designed {specifically} to simulate fluid flows. Built on a C++ platform, it offers both QC and CFD tools in one place. With QFlowS
we implement a modified version of the class of algorithms, now termed Quantum Linear Systems Algorithms (QLSA). Under some caveats, these algorithms promise to solve a linear system of equations given by the matrix inversion problem $A\mathbf{x}=\mathbf{b}$, with up to an exponential speedup compared to known classical algorithms. In recent years a number of efforts based on continuum methods using QLSA \cite{childs2021high,tosti2022review}, variational quantum algorithms \cite{leong2022variational,lubasch2020variational}, amplitude estimation methods \cite{gaitan2020finding,oz2023efficient}, {lattice based methods \cite{todorova2020quantum,itani2023quantum, budinski2021quantum}} and quantum-inspired methods \cite{gourianov2022quantum}, have been undertaken to solve linear and nonlinear PDEs. However most of these efforts have been theoretical, lacking gate-level quantum numerical simulations and analysis of the resulting flow field, or proper estimates of the actual errors involved. 

In particular, a full-gate level quantum simulation is implemented on QFlowS to solve the unsteady Poiseuille and Couette flow problems. We implement both the fundamental form of QLSA, called the Harrow-Hassidim-Lloyd (HHL) algorithm \cite{harrow2009quantum} and its more recent counterpart \cite{childs2017quantum}{,} based on the linear combination of unitaries (LCU). In addition, we prescribe suitable quantum state preparation protocols and propose a hitherto unseen quantum post-processing (QPP) protocol to compute {\it in situ} nonlinear functions of the resulting flow solution. In particular we obtain the viscous dissipation rate $\varepsilon=\nu(\frac{\partial u}{\partial y})^{2} $ 
averaged over the flow field $u$, $\nu$ being the viscosity.
Together, this forms an end-to-end implementation, which alleviates, to some extent, the restrictions posed by both quantum state preparation and the measurement of qubits---which are otherwise the major limiters of the theoretical quantum advantage \cite{alexeev2021quantum, harrow2009quantum,childs2021high,aaronson2015read}. Although the proposed algorithms are far from realistic fluid simulations, they make quantum implementations more amenable for small systems, and inform the running of codes on near-term NISQ machines while attempting to preserve the quantum advantage.

By necessity the paper uses a number of acronyms. For convenient reference, a glossary is provided in \textit {Appendix} \ref{sec:glossary}.

\begin{figure*}[htb!]
     
    \subfloat[]{
    \includegraphics[trim={2cm 14.4cm 2cm 12.5cm},clip=true,scale=1.1]{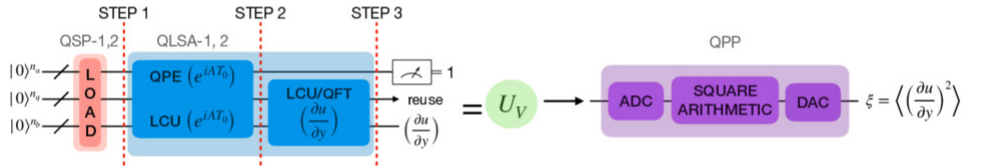}}\\
\subfloat[]{        
        \includegraphics[scale=0.21]{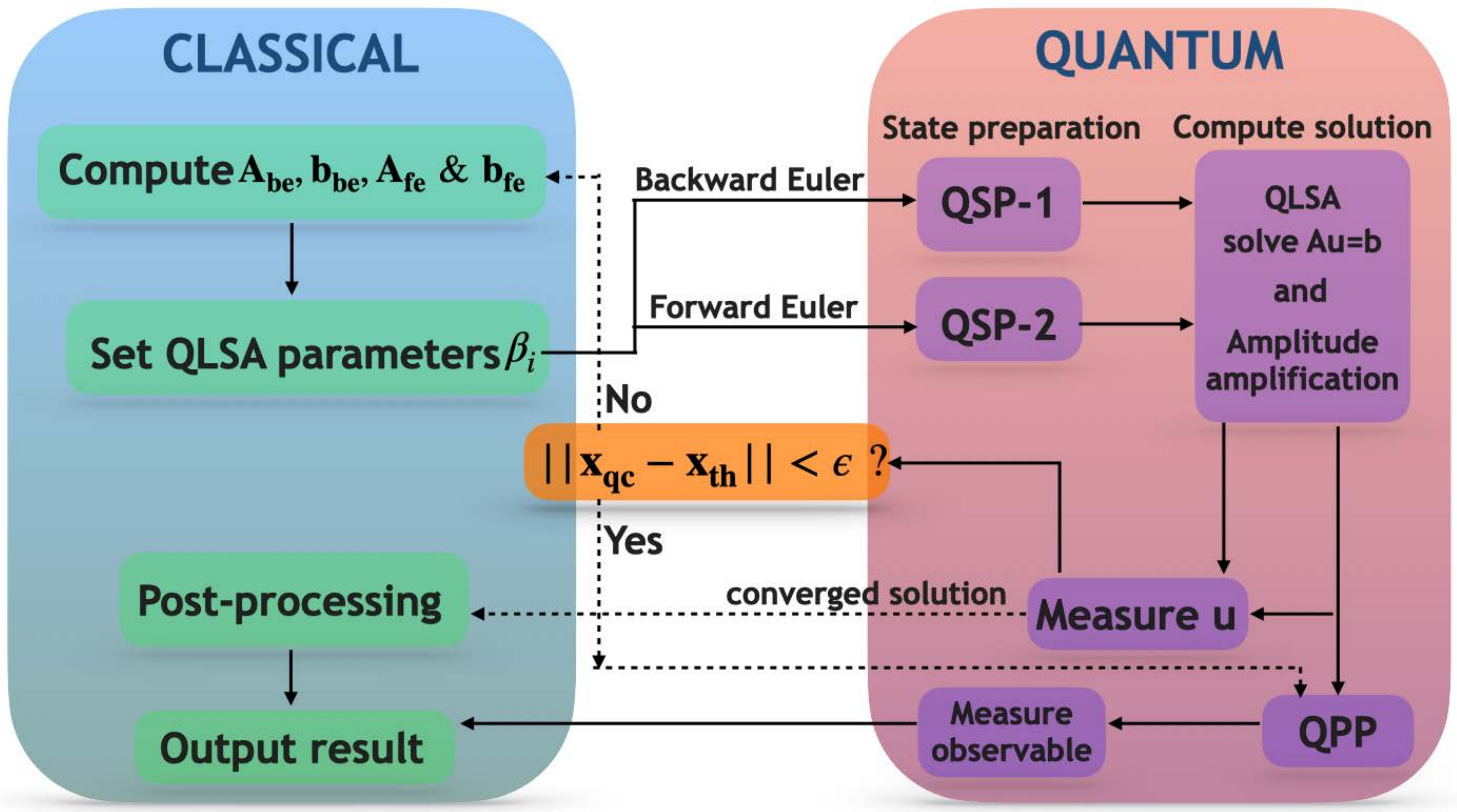}

         }
         \subfloat[ ]{
        \includegraphics[trim={0.6cm 0.1cm 1.3cm 0.3cm},clip=true,scale=0.4]{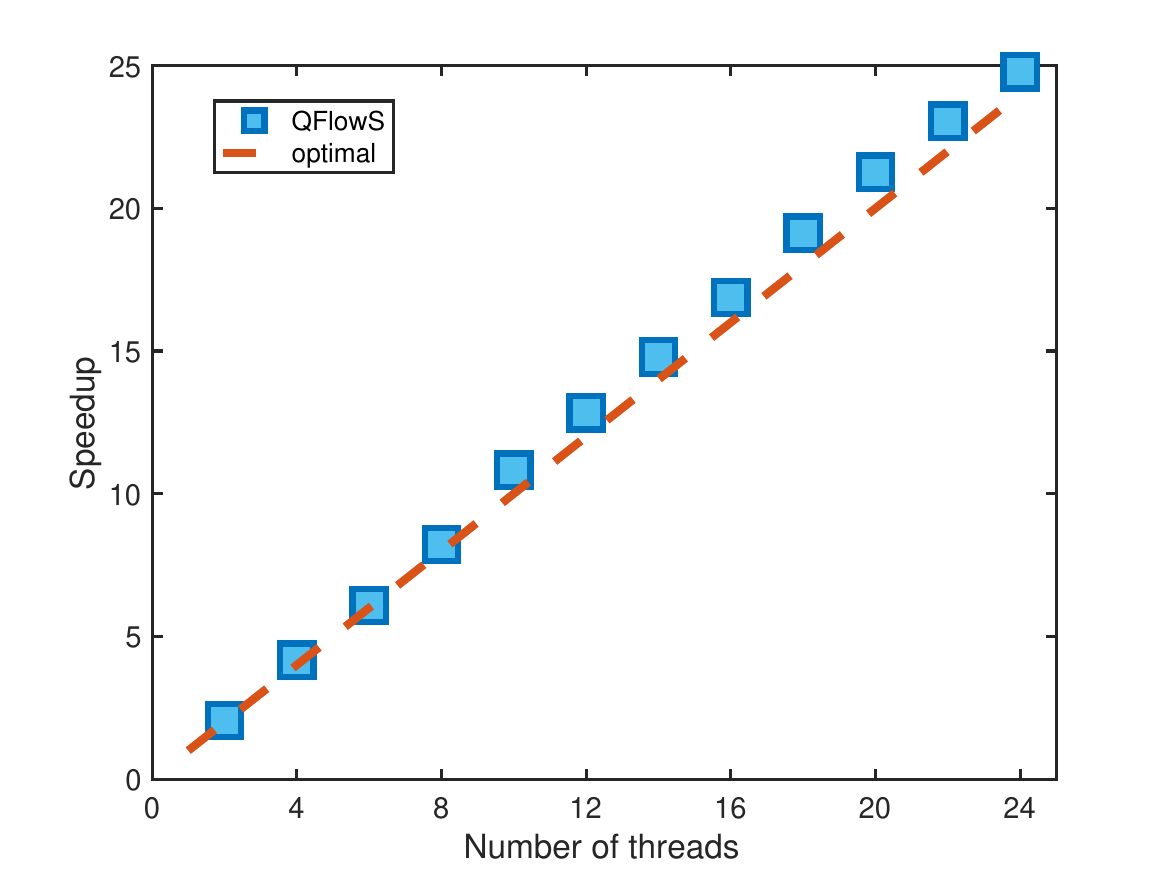}}
        
  \caption{\justifying (a) shows the modified QLSA circuit with QSP, forming an oracle $U_{V}$ that is employed recursively in the QPP protocol for computing viscous dissipation rate by a combination of Quantum Analog-Digital converters. (b) The working flowchart of the hybrid quantum-classical algorithm. (c) The strong scaling performance of QFlowS with super-linear speedup when run on single node, parallelized with OpenMP up to a total of 24 threads on NYU's Greene supercomputer. The performance is for simulations of a 20 qubit circuit of depth 422, performing a Quantum Fourier Transform (QFT) and inverse QFT algorithm. The standard deviations are smaller than the blue square symbols around the mean,computed over an ensemble of 8000 simulations with random initial quantum states.}
   \label{fig:Hybrid setup}
\end{figure*}

\section{Linear flow problems}
\label{sec:Lin flows}
We consider the well known 1D unsteady Poiseuille and Couette flows (schematic shown in \textit{SI Appendix} figure \ref{fig:flow setup} that are linear dissipative flows which describe for instance, micro-channel flows (e.g., in micro-chips, blood capillaries and syringes) or lubricant flows around bearings. The framework outlined in this work is readily {extendable} to the linear advection-diffusion with constant advection velocity. More generally, this algorithm caters to the class of elliptic and parabolic PDEs described by {d-dimensional} Laplace, Poisson and heat equations. Under certain boundary conditions, the flows under discussion admit exact analytical solutions, thus making them ideal candidates for evaluating the performance of the quantum solver. Some earlier works such as \cite{Cao_2013,montanaro2016quantum} made some important observations in possible implementations on QC for similar problems and estimated theoretical upper bounds of their complexities. The general form of the governing PDEs considered here (assuming no body forces or source terms) are given by the momentum and mass conservation relations 
\begin{equation}
        \frac{\partial \textbf{u}}{\partial t} + \mathbf{C}\cdot\nabla \mathbf{u} = \frac{1}{Re} \nabla ^{2} \textbf{u} -  \nabla \textbf{p}\label{eq:governing1} 
\end{equation}
\begin{equation}
    \nabla \cdot \textbf{u} = 0 .
    \label{eq:incompressibility}
\end{equation}
where $\mathbf{u} = (u,v,w)$ is the velocity field, $\mathbf{C}$ is a constant advection velocity (which is zero in the fully-developed state of flows considered here), $p$ is pressure field, $Re = UD/\nu$ is the Reynolds number, $U$ is the characteristic velocity, $\nu$ is the kinematic viscosity and $D$ is the separation between the boundaries. 
The so-called fully developed 1D case (applicable for all discussions from here on) reduces to \begin{equation}
    \frac{\partial u}{\partial t} = \frac{1}{Re} \frac{\partial^{2} u}{\partial y^{2}} -  \frac{\partial p}{\partial x},
    \label{eq:poiseuille flow}
\end{equation}where the velocity varies only along y (wall-normal direction), and the pressure gradient $\frac{\partial p}{\partial x}$ is a constant. The boundary conditions are no-slip with $u(0,t) = u(D,t) = 0$ for the Poiseuille flow and $u(0,t) = 0$ and $u(D,t) = 1$ for the Couette flow. The initial condition for the temporal evolution is set to be a uniform flow $u(y,0)= u_{in} = 1$. 

We reiterate that this problem is simple from the standpoint of the sophisticated advances of classical CFD. Nevertheless, this is an excellent starting point for demonstrating the viability of quantum algorithms for CFD---this being the spirit of the present work.

\subsection{Hybrid quantum-classical numerical setup}
\label{subsec:hybrid setup}
The goal now is to solve eq. (\ref{eq:poiseuille flow}) by means of QLSA, which thus necessitates eq. (\ref{eq:poiseuille flow}) to be recast as a linear system of equations. To do this, we consider the method of finite differences to discretize the computational domain in both space and time. Details of these schemes, their stability considerations and the resulting matrix equations that form the input to the quantum algorithm is  outlined in \textit{SI Appendix} \ref{subsec:FDM}. The well-known second order central difference scheme is used to discretize the Laplacian operator for $N_{g}$ grid points, while both forward and backward Euler (FE and BE from here on) schemes are implemented to discretize {time. This procedure yields a} set of three possible matrix equations $\mathbf{A}_{be1}\bar{u} = \mathbf{b}_{be1}$, $\mathbf{A}_{be2}\bar{u} = \mathbf{b}_{be2}$ and $\mathbf{A}_{fe}\tilde{u} = \mathbf{b}_{fe}$, {where be1 corresponds to the iterative backward-Euler scheme (system solved for every time step iteratively), and be2 and FE correspond to the one-shot backward Euler and forward Euler schemes (system solved for all time steps in one shot), respectively.}  

To solve these equations, a hybrid quantum-classical method is developed (schematic flow chart is shown in figure \ref{fig:Hybrid setup}(b)). 
The preconditioning and computations of the elements of the matrices $\mathbf{A}_{be1}, \mathbf{A}_{be2},\mathbf{A}_{fe}$ and vectors $\mathbf{b}_{be1},\mathbf{b}_{be2},\mathbf{b}_{fe}$ are done classically. Certain parameters required (as elucidated later) for quantum state preparation (e.g., rotation angles and decision trees) and for Hamiltonian simulation (time $T^{*}_{0}$) (collectively termed $\beta$)  are pre-computed classically as well. $N$ from here on refers to the dimension of final matrix system that results from these considerations. With this on hand, the inputs are first loaded on the QC by the quantum state preparation algorithms (QSP-1,2) and the resulting linear system of equations is then solved by QLSA. In the case of iterative BE, $\mathbf{A}_{be1}\bar{u} = \mathbf{b}_{be1}$ is solved for velocities $\bar{u}$, at every time step until convergence (residue reaching a tolerance $\leq \epsilon_{tol} = 10^{-6}$), which is checked classically. In a contrasting setup, BE and FE are used to set up, respectively, $\mathbf{A}_{be2}\tilde{u} = \mathbf{b}_{be2}$ and $\mathbf{A}_{fe}\tilde{u} = \mathbf{b}_{fe}$, giving $\tilde{u} = [u(y,0),u(y,dt),\cdots,u(y,T)]$, in one shot, $\forall t\in[0,T]$. It is important to note that even the BE method can be {set up} such that the solution is computed for {all $t$ at once.} However, in the absence of efficient state preparation and measurement protocols, measuring the solution and re-preparing the state for the next time step are $\mathcal{O}(N_{g})$ operations that eliminate any quantum advantage, making the overall algorithm no better than classical solvers (and with additional errors due to quantum measurements). In any case, it is still worthwhile establishing how the method fares as a plausible alternative to classical simulations. 

The end solution is either: (a) simply read by quantum measurements for post-processing on a classical device, or (b) {post-process, in-situ, a quantum device using the QPP protocol introduced here.} The former, at the level of a simulator, allows one to validate the correctness of solutions and redesign the circuit as required. In the latter case, only a single target qubit and few ancillas are measured, {outputting} \textit{one} observable---which is a real-valued \textit{nonlinear} function of the velocity field. Apart from computing nonlinear functions, this circumvents expensive and noisy measurements of entire quantum states and more importantly preserves quantum advantage (to the extent possible). 
\section{Quantum Flow Simulator - QFlowS}
\label{sec:QFlowS}
In \cite{bharadwaj2020quantum,quantiki} several commercially available quantum simulation packages are listed. Most of them, for instance Qiskit (IBM), Quipper \cite{green2013quipper} and QuEST \cite{jones2019quest}, are constructed for general purpose quantum simulations and are highly optimized for such operations, making it hard to customize the fundamental subroutines and data structures for CFD calculations. On the other hand, there are softwares such as ANSYS and OpenFOAM that perform solely classical CFD simulations. With the motivation of having a single bespoke quantum simulator for CFD, we unveil here a high performance, gate level quantum-simulation toolkit, which we call \textit{QFlowS}; it is based on a C++ core and designed to be used both independently or as part of other software packages. It has a current capability of 30+ qubit simulation of custom quantum circuits. It also has several built-in gates and quantum circuits that could be used readily, while also being able to probe different quantum state metrics (such as the norm, density matrix and entanglement). Along with these, it includes basic CFD tools needed to set up flow problems making it versatile for QCFD simulations. Noise modelling is in progress and forms the major part of future software development. QFlowS is also being increasingly parallelized for optimal performance on supercomputers. For instance, figure \ref{fig:Hybrid setup}(c) shows the strong scaling performance using OpenMP. The performance is measured while running on NYU's Greene supercomputing facility. On a single medium-memory computer node (48 cores: 2x Intel Xeon Platinum 8268 24C 205W 2.9GHz Processor) and for a choice of 20 qubits, we measure the run-time (by omitting the one-time initial overhead processes) of a QFT-IQFT circuit action on an ensemble of randomly initialized quantum states. We observe near optimal and at times super-optimal scaling with increasing number of threads up to 24. Super-optimality arises when the quantum circuit is sparse, causing lesser quantum entanglement. Every single circuit layer operation is distributed over many worker threads, whose cache size exceeds the size of quantum state subspace being handled, thus making them closely parallel. \textit{SI Appendix} \ref{sec:QFlowS overview}, summarizes features of QFlowS.

\section{Quantum Linear Systems Algorithm (QLSA)} 
\label{sec:QLSA}
One of the first quantum protocols for solving equations of the form $A\vec{x}=\vec{b}$ is the HHL algorithm \cite{harrow2009quantum}, {which is the basis for what we refer to here as} \textbf{QLSA-1}. In \cite{harrow2009quantum} it was shown that: \textit{For a hermitian and non-singular matrix A $\in \mathbb{C}^{2^{n}\times2^{n}}$, vector b $\in \mathbb{C}^{2^{n}}$ ($N=2^{n}$), given oracles to prepare A and b in $\mathcal{O}(polylog(N))$, and a prescribed precision of $\epsilon > 0$, there exists an algorithm that computes a solution $x$ such that $|||x\rangle - |A{^{-1}}b\rangle||\leq \epsilon $ in $\mathcal{O}(polylog(N)s^{2}\kappa^{2} /\epsilon)$}, where $\kappa$ is the condition number of the matrix and $s$ is the sparsity. This shows that the algorithm is exponentially faster than classical alternatives, but there are important caveats \cite{aaronson2015read}). Later works \cite{vazquez2021efficient,giovannetti2008architectures} attempted to address these caveats, while some others \cite{georgescu2014quantum,berry2014exponential,berry2015hamiltonian} fundamentally improved the method by reducing error complexity from $poly(1/\epsilon)$ to $poly(\log(1/\epsilon))$. Consequently refs.~\cite{ambainis2010variable,berry2014exponential,childs2021high} led to a more precise class of QLSA methods based on the linear combination of unitaries (LCU) \cite{childs2017quantum} which we shall refer to as \textbf{QLSA-2}. In \cite{childs2017quantum} it was shown that under similar caveats of QLSA-1, we have: \textit{For a hermitian and invertible matrix A $\in \mathbb{C}^{2^{n}\times2^{n}}$, vector b $\in \mathbb{C}^{2^{n}}$ , given oracles to prepare A and b in $\mathcal{O}(polylog(N))$, and a prescribed precision of $\epsilon > 0$, there exists an algorithm that computes a solution $x$ such that $|||x\rangle - |A{^{-1}}b\rangle||\leq \epsilon $ in $\mathcal{O}(polylog(N/\epsilon)\kappa )$}. 
This work implements {modified algorithms derived from} both these methods \cite{harrow2009quantum,childs2017quantum} by developing an efficient LCU decomposition strategy. 
In the QCFD context, we now explore methods suitable for preparing $\{\mathbf{b}_{be1},\mathbf{b}_{be2},\mathbf{b}_{fe}\}$ and the matrices $\{\mathbf{A}_{be1},\mathbf{A}_{be2},\mathbf{A}_{fe}\}$, to enable the post-processing of the solution $\tilde{u}$, in order to construct an end-to-end method.
\subsection{Quantum State Preparation}
\label{subsec:QSP}
To prepare quantum states that encode $\mathbf{b}_{be1}$, $\mathbf{b}_{be2}$ and $\mathbf{b}_{fe}$, we implement two different methods, both offering sub-exponential circuit depth complexity:

{\bf (QSP-1)} In the case of iterative BE, the vector $\mathbf{b}_{be1}$, prepared at every time step, is generally fully dense with sparsity $s_{b} \sim \mathcal{O}(N_{be1})$. In the specific cases of Poiseuille and Couette flows, and for the specific initial conditions considered here, the state prepared at every time step forms a discrete log-concave distribution (i.e., $\frac{\partial^{2}\log(b)}{\partial y^{2}}<0$ for $\forall t \geq 0$ ), which could also be confirmed from the analytical solution given by eq. (\ref{eq:analytical}) known for this case as \begin{align}
    u(y,t) = \sum_{k=1}^{\infty}&\Bigg[\frac{2(1-(-1)^{k})}{k\pi}\Big(1+\frac{\partial p}{\partial x}\frac{Re}{(k\pi)^{2}}\Big)\label{eq:analytical}\\ \nonumber &\sin\Big(\frac{k\pi y}{D}\Big)e^{\frac{-t}{Re}(\frac{k\pi}{D})^{2}}\Bigg] - \frac{Re}{2}\frac{\partial p}{\partial x}y(1-y).
\end{align} Even if the exact solution is not known, provided the initial condition is the only state preparation involved in the algorithm, flexibility exists for most flow simulations in choosing initial conditions that are log-concave.   Consequently one could invoke a \textit{Grover-Rudolph} state preparation \cite{grover2002creating} technique (or its more evolved {off-spring} \cite{rattew2022preparing,vazquez2022enhancing}) to offer an efficient way to encode data.  Two comments are useful. (i) Though this method could be used for arbitrary state-vectors (at the cost of exponential circuit depth), for an efficient state preparation, some information on the functional form of the state needs to be known \textit{a priori}---from analytical solutions, classical CFD, or by the measurement of the quantum circuit at intermittent time-steps, peeking into its instantaneous functional form. Here, we implement a similar method, which we shall refer to as \textbf{QSP-1}, based on \cite{grover2002creating,prakash2014quantum}, where it was shown that: \textit{Given a vector $\mathbf{b}_{be1} \in \mathbb{R}^{N_{be1}}$,state $\mathbf{b'}_{be1}$ can be prepared such that, $||\mathbf{b}_{be1}\rangle - |\mathbf{b'}_{be1}\rangle| < \mathcal{O}(1/ {poly}(N_{be1}))$ in $\mathcal{O}(\log(N_{be1}))$ steps.} (ii) Measuring all qubits of the register ($\sim \mathcal{O}(N_{be1})$) at every time step compromises the exponential speed-up and could introduce measurement errors. However, such a method of recursive state preparation and measurement could still prove to be useful with quantum advantage for a very small number of qubits \cite{pfeffer2022hybrid}. In any case, we implement this method here to explore if such a BE scheme gives accurate results with or without quantum advantage. 

{\bf (QSP-2)} In the case of the one shot methods, an alternative quantum state preparation method can be considered since $\mathbf{b}_{fe}$ is generally larger in size $\sim \mathcal{O}((m+p)N_{g})$ (this discussion applies similarly to $\mathbf{b}_{be2}$; refer to \textit{SI Appendix} \ref{subsec:FDM} for definition of terms used here). When considered together with all other registers that are initially set to $|0\rangle$,  $\mathbf{b}_{fe}$ is a highly sparse state vector with $s_{b} \sim \mathcal{O}(N_{g}m)$. For such states, we implement a sparse state preparation protocol 
\cite{mozafari2022efficient}, which we shall refer to as \textbf{QSP-2}. It provides an optimal circuit depth that scales only polynomially with vector size. This method involves constructing decision trees forming an alternative way to represent quantum states. Careful optimization on the structure of these trees leads to efficient state preparation whose complexity depends on the number of continuous pathways in the resulting tree structure. Thus, rephrasing here the result in \cite{mozafari2022efficient}, we have: Given an $n$-qubit initial state of size $N=2^{n}$, all set to $|0\rangle$, except for a sparse vector subspace $\mathbf{b}_{fe} \in \mathbb{R}^{N_{fe}}~(N_{fe} = \mathcal{O}((m+p)N_{g}))$, with sparsity $s_{b}= m \ll N$, then with only single qubit and CNOT gates, one can prepare such a state with in $\mathcal{O}(2kn)$ time, $k\times\mathcal{O}(n)$ CNOT gates and using 1 ancillary qubit, where $k(\leq m$) is the number of branch paths of the decision tree. 

Both QSP-1 \& 2 are elucidated with examples in {\textit{SI Appendix}} \ref{sec:QSP Append}.

\begin{figure*}[htb!]
     \centering{
    \subfloat[]{
    \includegraphics[trim={0.9cm 0.4cm 1.5cm 1cm},clip=true,scale=0.4]{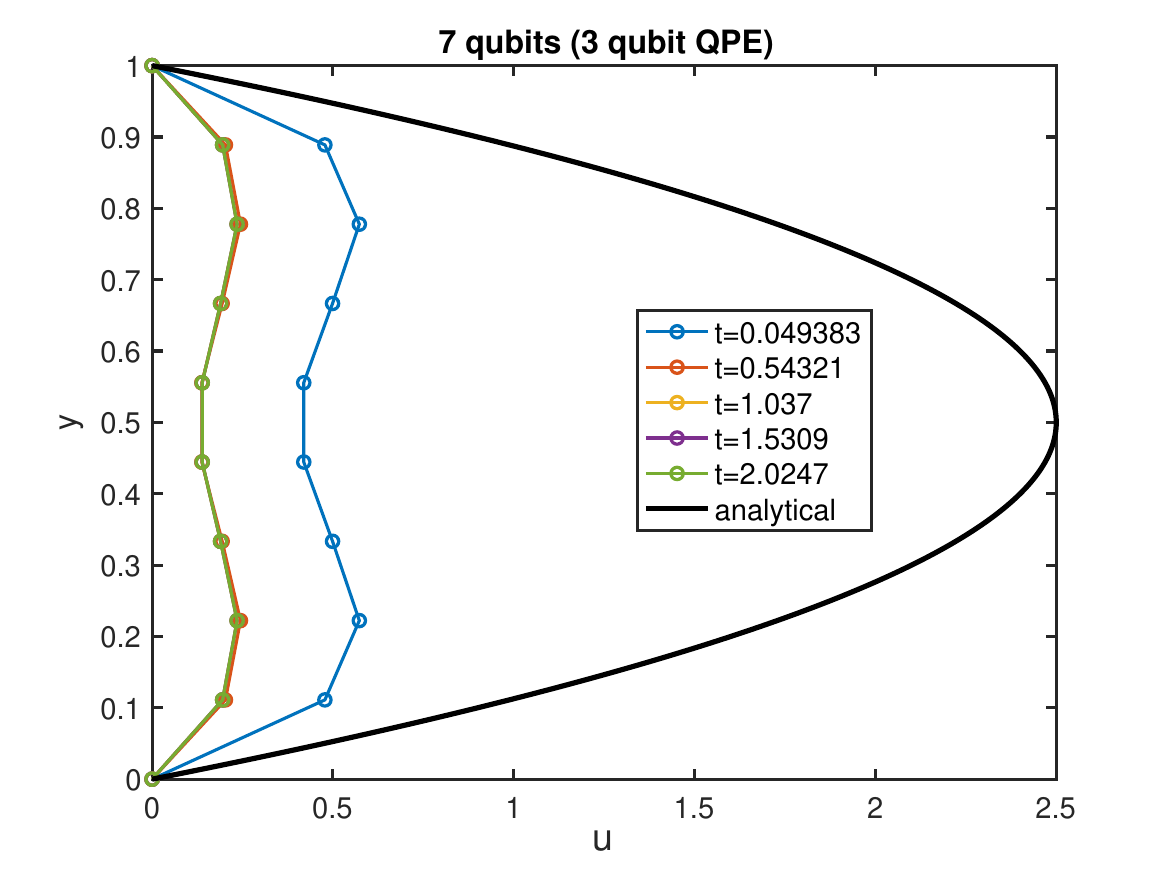}  }
    \subfloat[]{
         \includegraphics[trim={0.9cm 0.4cm 1.7cm 0.9cm},clip=true,scale=0.4]{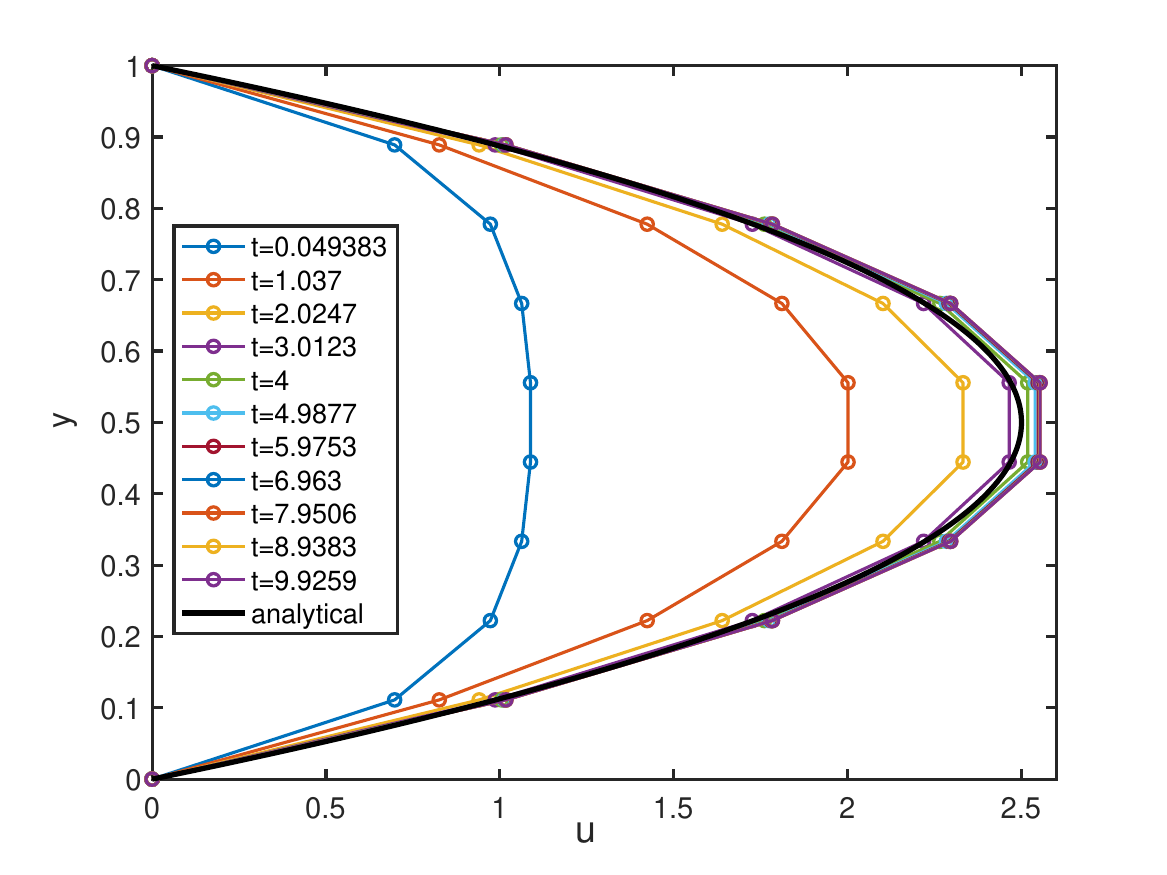}
    }
    
     \subfloat[]{
         \includegraphics[trim={0.9cm 0.2cm 1.5cm 0.9cm},clip=true,scale=0.4]{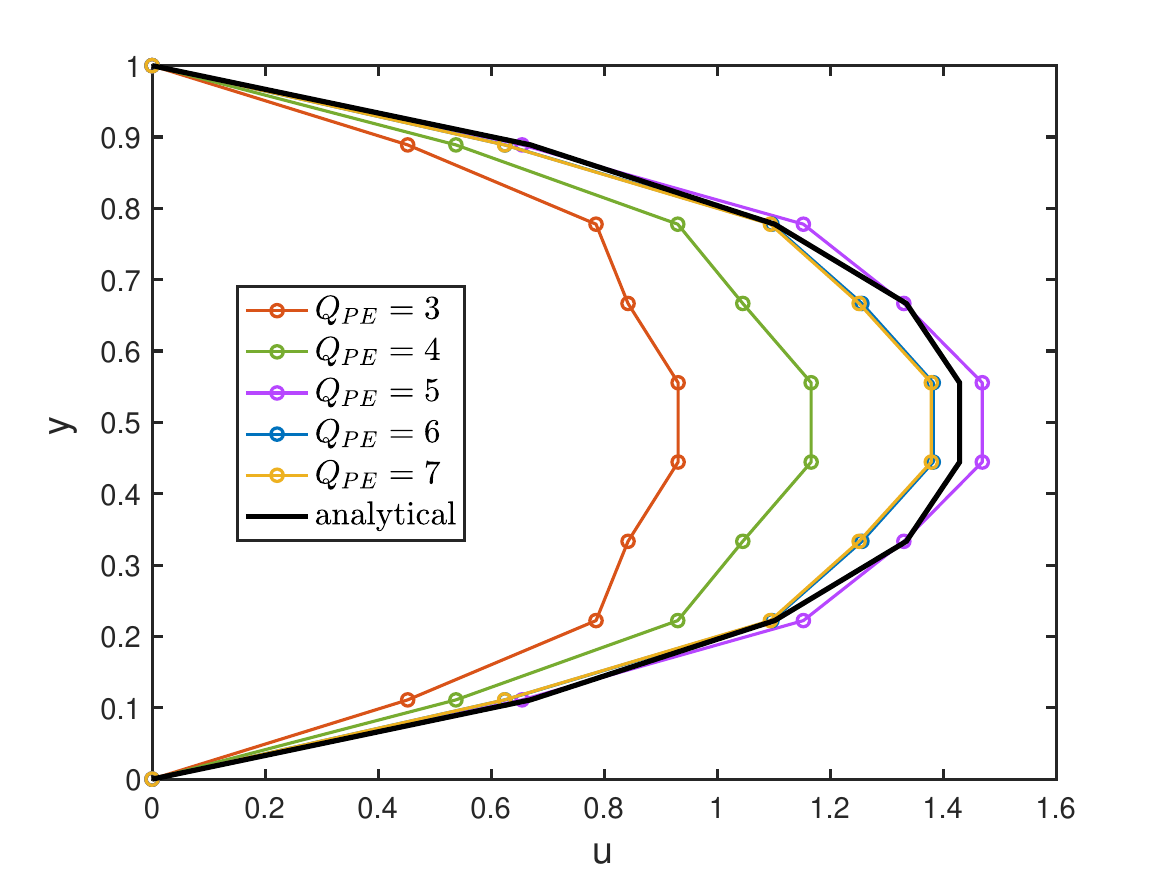}
    }
    \subfloat[]{
         \includegraphics[trim={0.9cm 0.2cm 1.5cm 0.9cm},clip=true,scale=0.4]{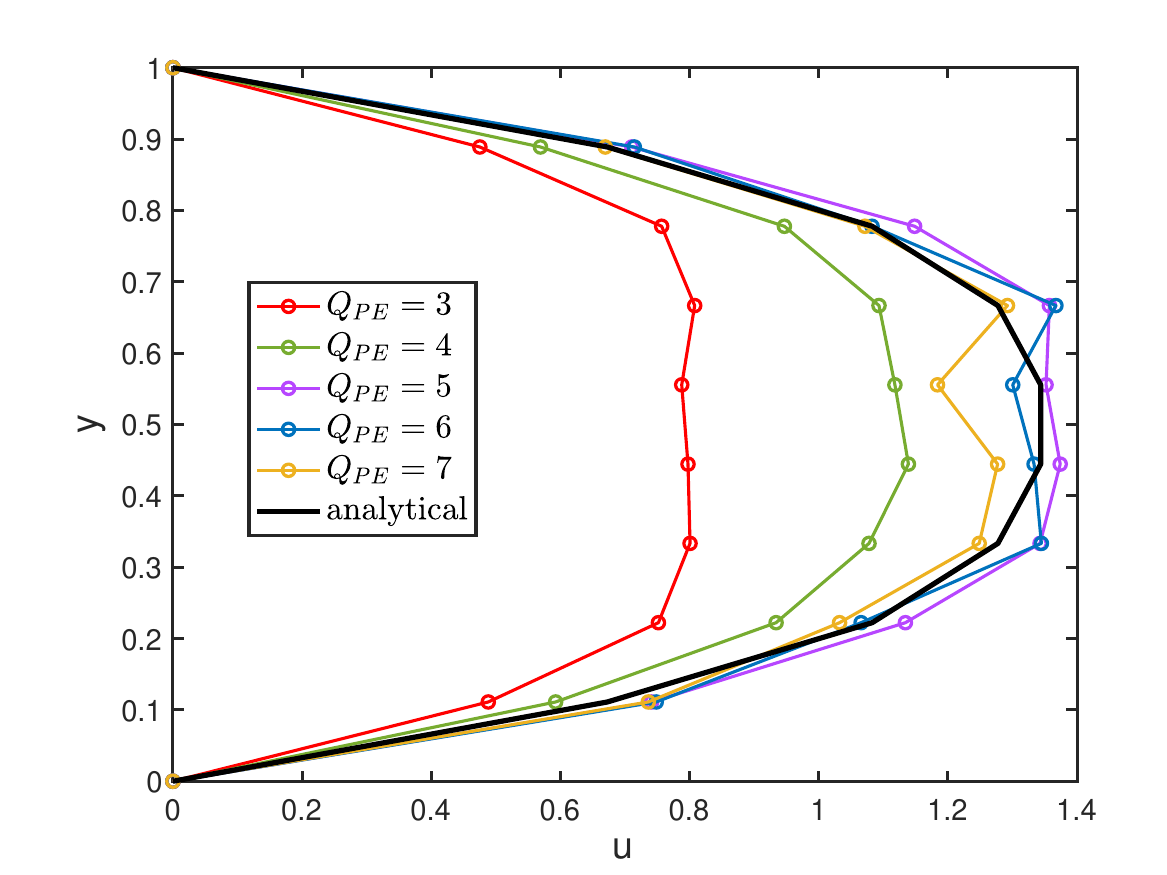}
    }}
    \caption{ \justifying (a) and (b) show the quantum simulation of the flow field evolving forward in time towards steady state (analytical parabolic solution shown as solid black line) using BE scheme that uses 7 and 14 qubits (3 and 10 QPE qubits ($Q_{PE}$), respectively), for $N=10$, $Re = 10$, $\partial p /\partial x = -2$ and $dt = 0.01$. The accuracy of the converged solution improves for higher $Q_{PE}$. Here the velocity field is solved for at every time step; (c) shows increasingly accurate converged solutions with increasing $Q_{PE}\in\{3,4,5,6,7\}$, but solved using the FE scheme where the velocity field is solved for all time steps in one shot, and only the final solution is extracted. Here $\alpha = 0.5$ is set to meet the the von Neumann stability criterion and the parameter $T^{*}_{0} = 5.0$ is fixed. (d) also shows the one shot method, solved with a BE scheme {and $T^{*}_{0} = 8.5$.} Though $\alpha = 0.5$ is set to maintain the same time step size as (c) for comparison), there is no hard constraint given that the method is unconditionally stable. }
    \label{fig:velocity field}
    \end{figure*}
 \begin{figure*}[htb!]
    \centering{
    \subfloat[]{
         \includegraphics[trim={0.4cm 0.4cm 1.5cm 0.9cm},clip=true,scale=0.4]{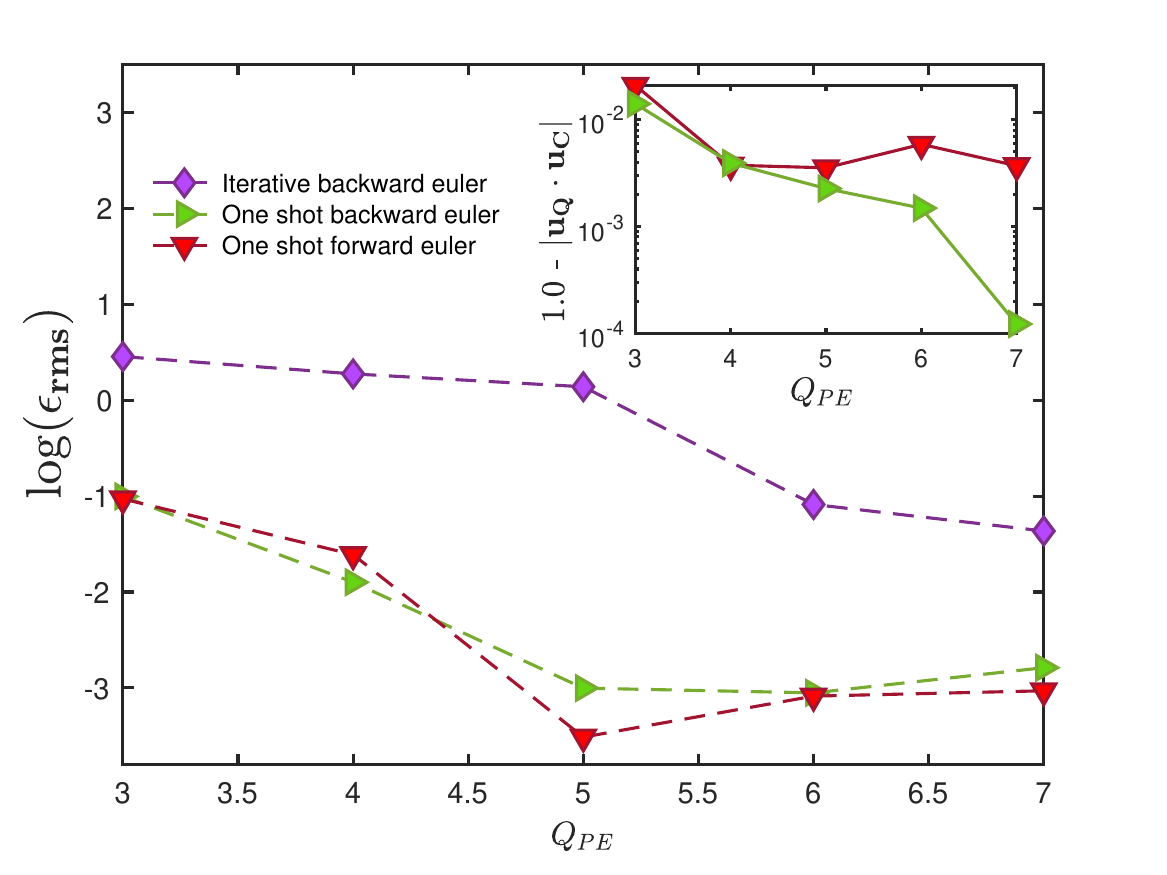}
    }
    \subfloat[]{ \includegraphics[trim={0.7cm 0.2cm 1.5cm 0.9cm},clip=true,scale=0.4]{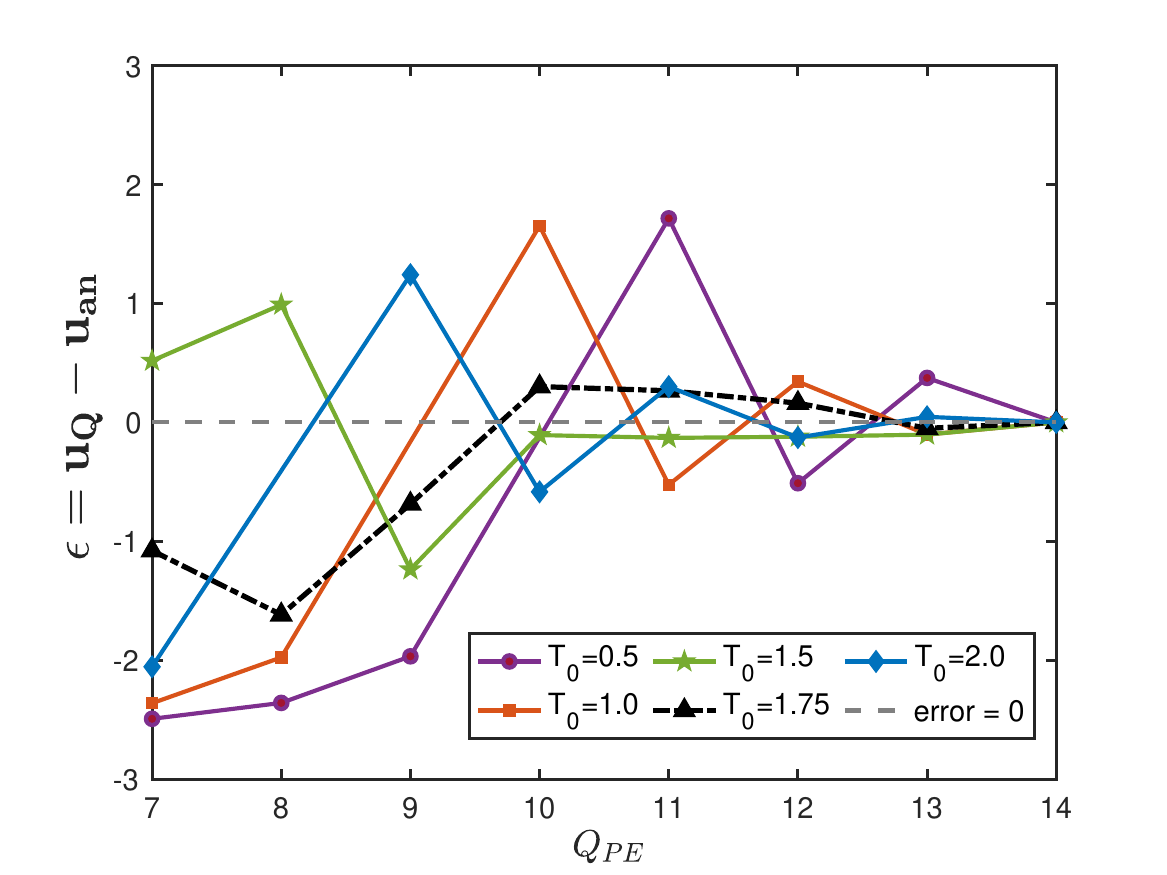}   
         }

    \subfloat[]{\includegraphics[trim={0.5cm 0.1cm 1.4cm 0.28cm},clip=true,scale=0.41]{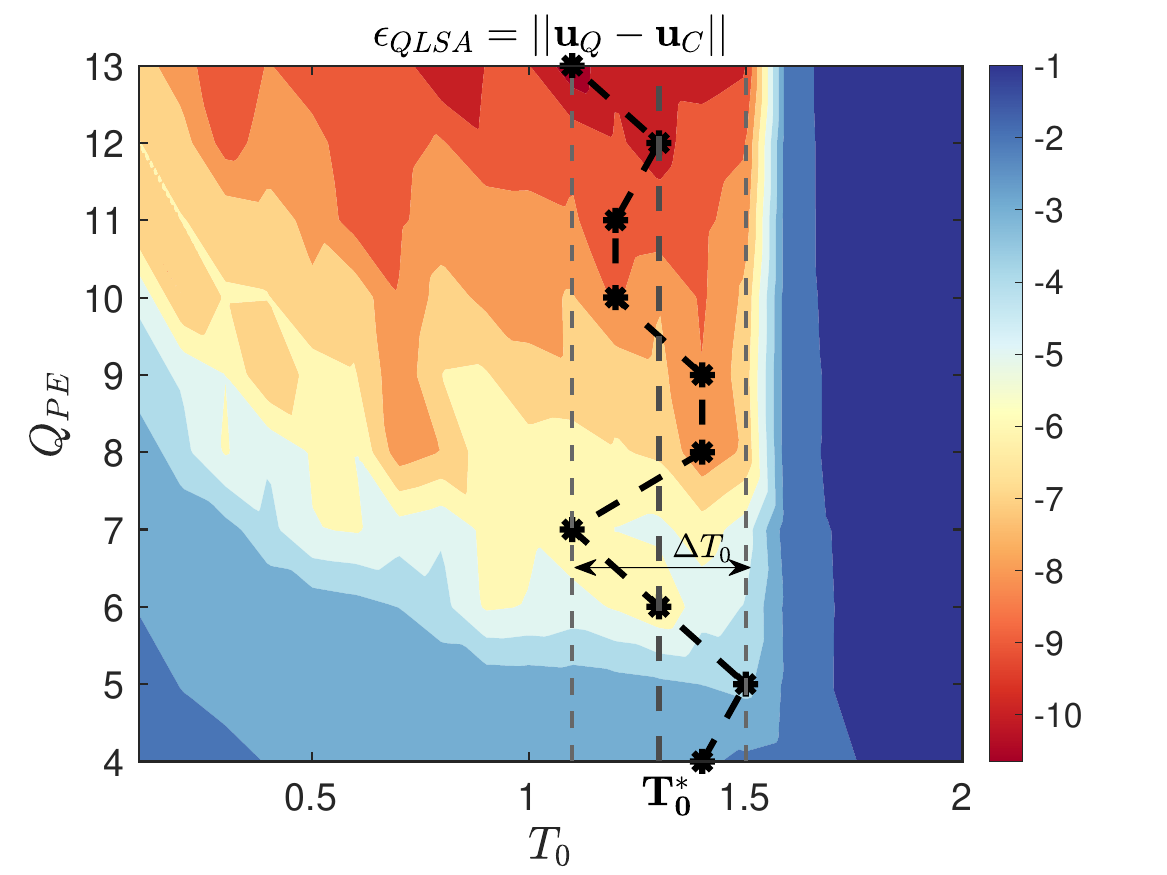}   
         }
    \subfloat[]{ \includegraphics[trim={0.6cm 0.1cm 1.8cm 0.8cm},clip=true,scale=0.41]{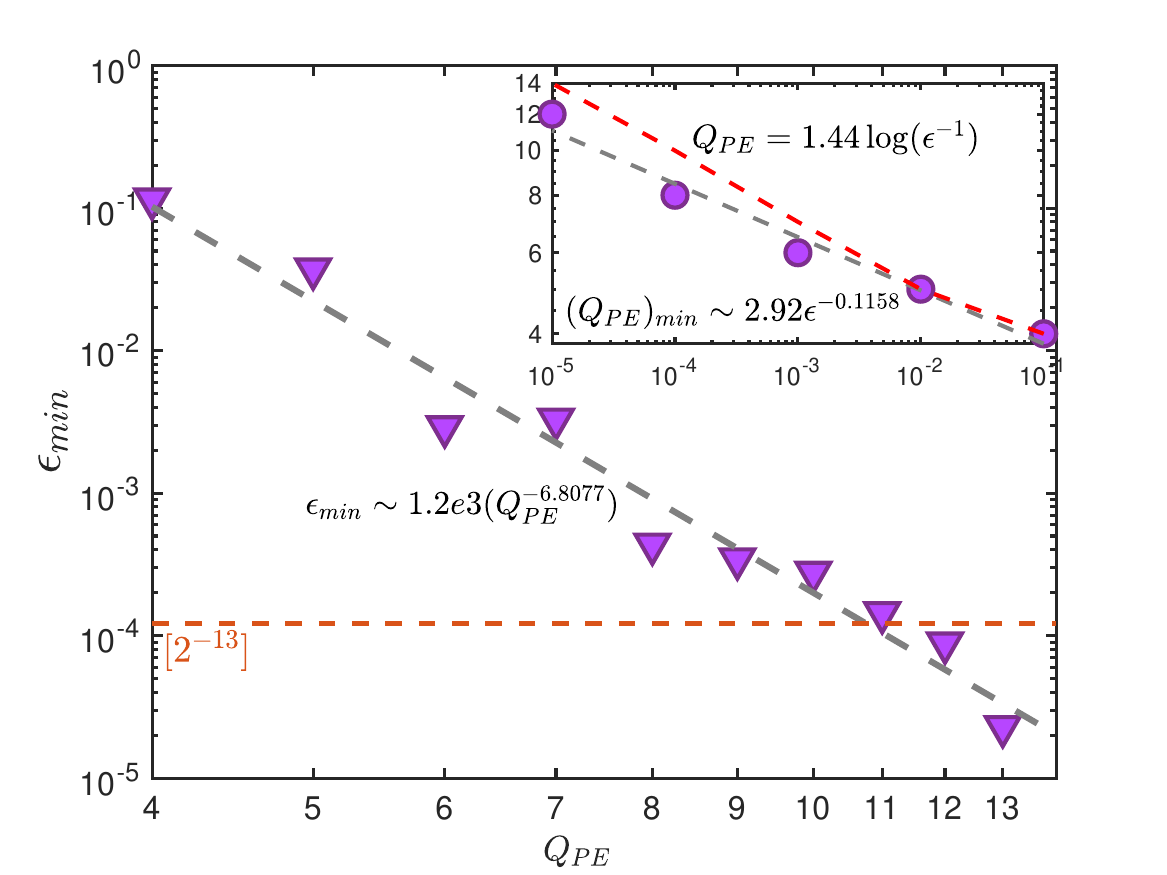}   
}}

\caption{(a) \justifying shows the absolute root-mean-square (RMS) error for the BE and FE cases shown in \ref{fig:velocity field}((a)-(d) against QPE qubits of both the step-by-step and one shot methods. To compare directly the one shot BE and FE methods, we show in the inset the quantity (1-fidelity) as a function of $Q_{PE}$. (b) shows the absolute error computed with respect to the analytical solution as a function of $Q$ for varying $T_{0}$. The black dotted line joining upright triangles shows the case for which the error magnitude and oscillation around the zero line are the least. (c) shows the contour plot of the QLSA {r}oot-mean-square error $\epsilon_{QLSA}$ computed with respect to the classical inversion solution for different $TQ = (Q_{PE},T_{0})$ pairs. The dotted black line traces the locus of the least/minimum error for each $(T,Q)$ pair in that range, which shows an oscillation around a unique median value $T^{*}_{0}$ specific to a given matrix (or $\kappa$). The color code is given by the thin vertical bar to the right. (d) shows the decay in the $(\epsilon_{QLSA})_{min}$, extracted from (c) with respect to only $Q_{PE}$ for a fixed $\kappa=18.8795$ and $T^{*}_{0} =1.3 $. The inset plots the $Q_{PE}$ (ordinate) required to achieve a specified $\epsilon_{QLSA}$ (abscissa).
 }
  
\label{fig:error}    
\end{figure*}

\section{Flow simulation results} 
\label{sec:Results}
We construct and solve the system given by eqs.\ (\ref{eq:poiseuille flow}) for $N_{g}=10$ and $Re = 10$.  We observe that the quantum solutions for the velocity field capture the physics both qualitatively and quantitatively. To discuss closely the utility of QFlowS we consider results from QLSA-1. As shown in figure \ref{fig:velocity field}(a,b) the converged steady state solution (using iterative BE) undershoots the analytic solution for 7 qubits, and performs better with a higher number of qubits ($Q_{PE} \geq$ 12){. These qubits refer to the ones} allocated for the Quantum Phase Estimation (QPE) algorithm, which in turn decides the quantum numerical precision; {$Q$ is the total number of qubits in the circuit.} Similarly, the converged solution for the one shot FE and BE cases also become more accurate with respect to both analytical and classical CFD solutions, with increasing number of qubits, as seen in figure \ref{fig:velocity field}(c) and (d) respectively. 

Our experience is that, between the three schemes, the one shot FE and BE turns out to be more accurate than iterative BE for increasing number of qubits, as shown in figure \ref{fig:error}(a), where the error $\epsilon_{rms}$ is computed with respect to the analytical solution. Among the two, one shot schemes, though quantitatively their error behavior is nearly the same, (i) We see some spurious oscillation like error in the velocity profile for the BE case as seen in figure \ref{fig:velocity field}(d) for higher $Q_{PE}$. (ii) The BE case, however, has no stability based restrictions as the FE case making it more flexible on the choice of $dt$. (iii) When accuracy in temporal discretization is of the concern the one shot FE fares better. The performance can also be measured by computing fidelity{, which quantifies the degree of overlap of the quantum solution with respect to the classical ($\vert\mathbf{u_{Q}\cdot u_{C}} \vert \leq 1$.) This is plotted} in the inset of figure \ref{fig:error}(a), which shows the BE to perform better than FE. However fidelity might not always be a good indicator to performance as illustrated in \textit{SI Appendix} \ref{subsec:Couette}. Both QLSA-1 and QLSA-2 rely on a variant of the phase estimation algorithm which contributes most to the total error QPE; in QLSA-2, the Gapped Phase Estimation (GPE) is rather computationally inexpensive and less erronous than in QLSA-1\cite{childs2021high}). 
In the case of phase estimation, the operator/matrix under consideration is exponentiated first as $e^{iAT_{0}}$, where $T_{0}$ is the Hamiltonian simulation time. An optimal choice $T^{*}_{0}$ (unknown a priori) arranges the eigenvalues $\lambda_{j}$ that are spectrally decomposed in the basis of $A$ as $\sum_{j}e^{i\lambda_{j}T^{*}_{0}}|u_{j}\rangle\langle u_{j}|$ {producing} the best {$Q_{PE}$}-bit binary representation $|\Tilde{\lambda}_{j}\rangle_{Q_{PE}}$. {This choice} also minimizes possible truncation errors and any spurious quantum numerical diffusion. 

In case of QLSA-1 it is important to note {that the smallest eigenvalue will contribute most to the error, which eventually focuses our interest in estimating $\lambda^{-1}$.} Therefore ensuring that the smallest value representable with $Q_{PE}$ qubits {(\textit{least count} of QPE)} $= 2^{-Q_{PE}}$ is $ \leq \tilde{\lambda}_{min}$ is essential. The error for all cases shown in figure \ref{fig:error}(a) has a gradual {step-like} decay because increasing $Q_{PE}$ in small steps (of $\mathcal{O}(1)$) does not lower the least count appreciably (in $\log_{10}$ or $\log_{e}$) as $Q_{PE}$ gets larger. In case of QLSA-2, though it {avoids} a full blown QPE, the right choice of $T_{0}$ (for the Fourier approach \cite{childs2021high}){, the LCU coefficients and basis, along with quantum amplitude amplification,} is still crucial for better accuracy.

{When we probed further at the level of the flow field}, the choice of $T_{0}$ seemed to exhibit non-trivial effects; for instance, in the iterative BE case, when gradually increasing $Q_{PE}$ qubits, the converged solution either undershot or overshot the analytical solution initially. This is captured in figure \ref{fig:error}(b), where the error $\epsilon$ (with respect to the analytical solution $\mathbf{u_{an}}$ of the center line velocity solution) oscillates around $\epsilon = 0$ before converging to it for $Q_{PE} > 12$. For the specific case shown here, a choice of $T^{*}_{0} = 1.75$ (dotted black line) has the least oscillation of the error and best accuracy. 

We can now ask: what combination $TQ = (T_{0},Q_{PE})$ gives the least error? To answer this better, we take a sample matrix equation system (of size $8\times8$ and $\kappa = 18.8795$) and solve it for different $TQ$. We then make a contour plot of the QLSA error $\epsilon{_{QLSA}}= ||u_{Q}-u_{C}||$, as shown in figure \ref{fig:error}(c) and trace the path of least error $\epsilon_{min}$ for each $TQ$. Further, the range of the $T_{0}$ scan can be reduced with some initial estimates to the lower and upper bounds of $\lambda_{min}$ and $\lambda_{max}$ \cite{wolkowicz1980more} such as, $\beta_{1} - \beta_{2}\sqrt{N-1}\leq\lambda_{min}\leq \beta_{1} - \beta_{2}/\sqrt{N-1}$ and $\beta_{1} + \beta_{2}/\sqrt{N-1}\leq\lambda_{max}\leq \beta_{1}+\beta_{2}\sqrt{N-1}$, where $\beta_{1} = \text{Tr}(A)/N$ and $\beta_{2} = (\text{Tr}(A^{2})/N - \beta^{2}_{1})^{1/2}$. We observe that the optimal $T^{*}_{0}$ for all combinations lies in a fairly small range $\Delta T_{0} \sim 0.1$. This is a unique value lying along the median of this range, $T^{*}_{0} \approx 1.3$ for which the system performs best. This means that {for such a $T^{*}_{0}$,} all or most eigenvalues are best represented in binary form with $Q_{{PE}}$ qubits (one or some of the eigenvalues could also turn out to be represented exactly). Further, given $T^{*}_{0}$, with increasing number of qubits, the minimum error exhibits a power law decay $\epsilon_{min} \sim Q_{PE}^{-6.81}$ as shown in figure \ref{fig:error}(d), reaching $\sim 10^{-5}$ at around 13 qubits. The exponent becomes increasingly negative with decreasing $\kappa$, since the range of eigenvalues becomes smaller and more eigenvalues tend to be easily representable with a given number of qubits. The thick horizontal red line shows the least count for $Q_{PE} = 13$; here, $\epsilon_{min} < 2^{-13}$. 
\begin{figure*}[htpb!]

    \subfloat[]{
        \includegraphics[trim={1cm 0cm 1.8cm 1.1cm},clip=true,scale=0.35]{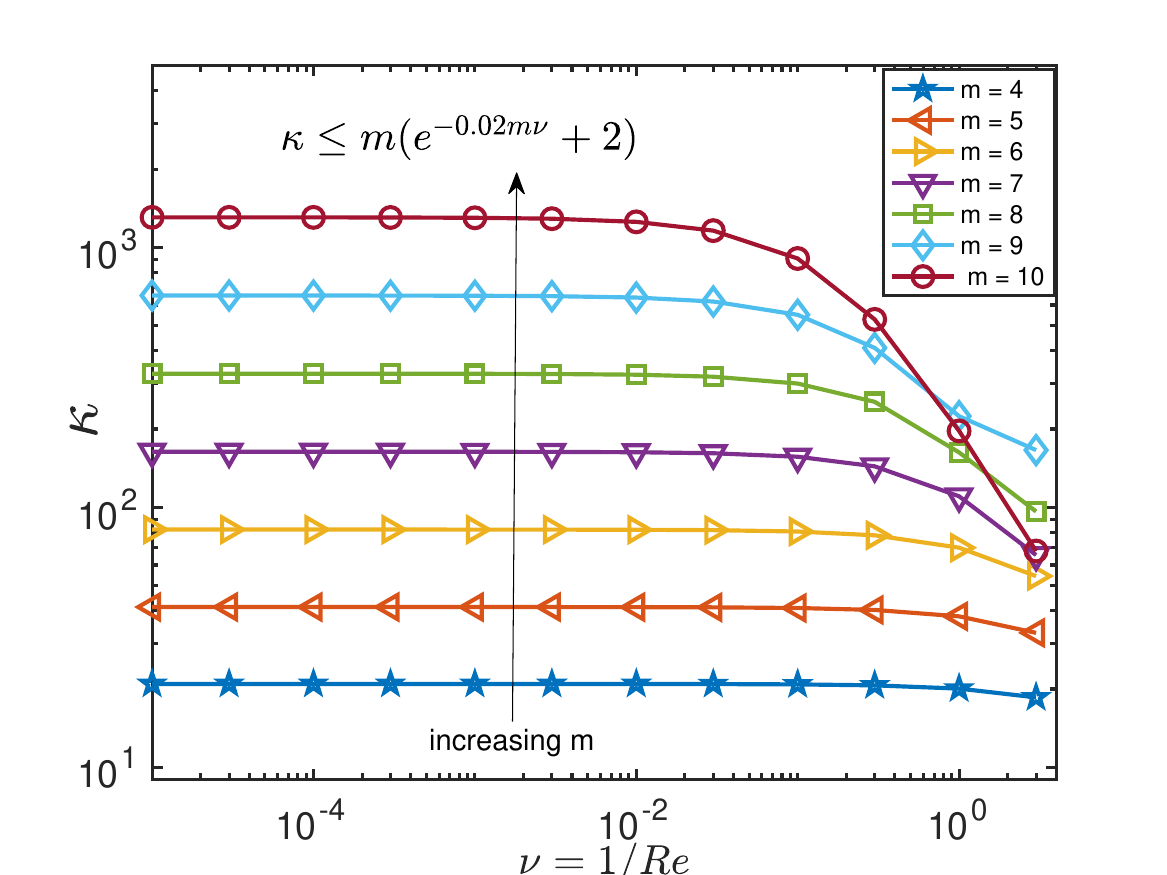}
    }
    \subfloat[]{
        \includegraphics[trim={0.5cm 0cm 1.8cm 1cm},clip=true,scale=0.35]{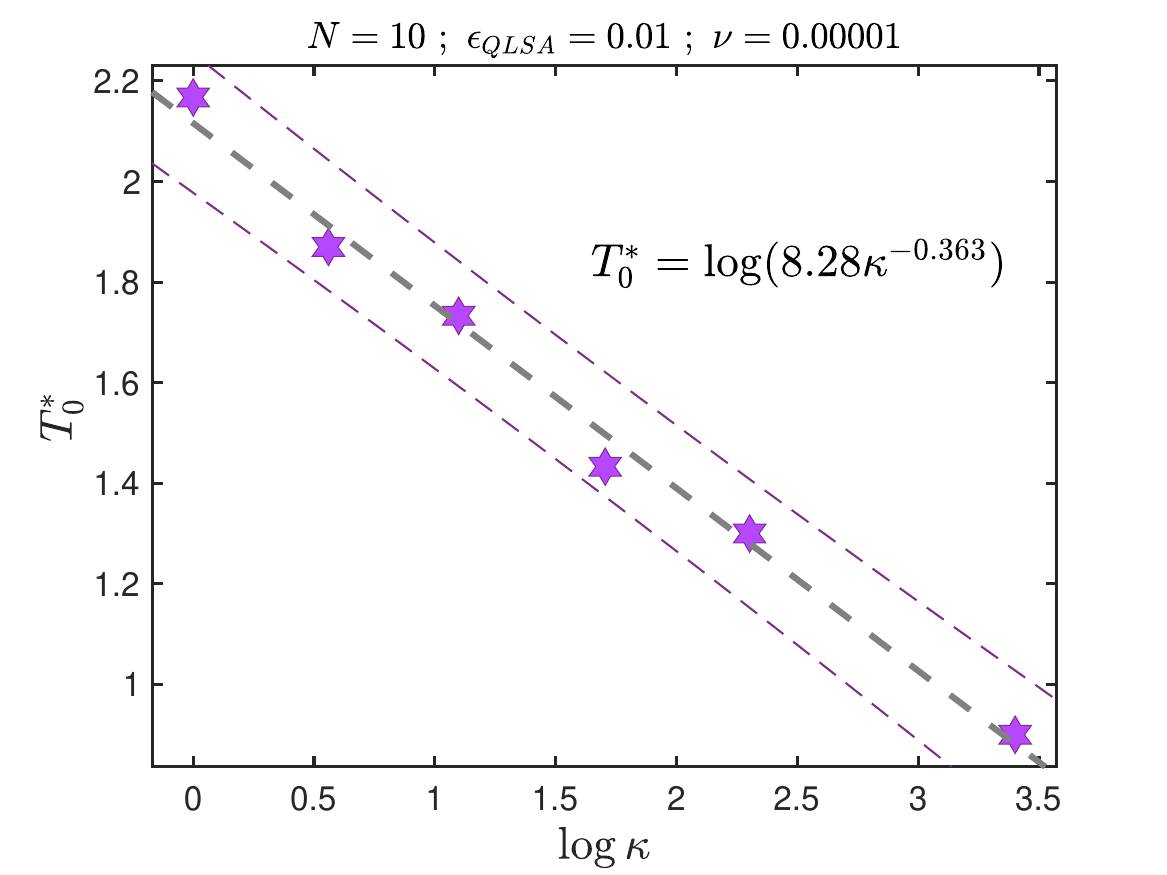}
    }
    \subfloat[]{
    \includegraphics[trim={0cm 6.3cm 1cm 7.5cm},clip=true,scale=0.31]{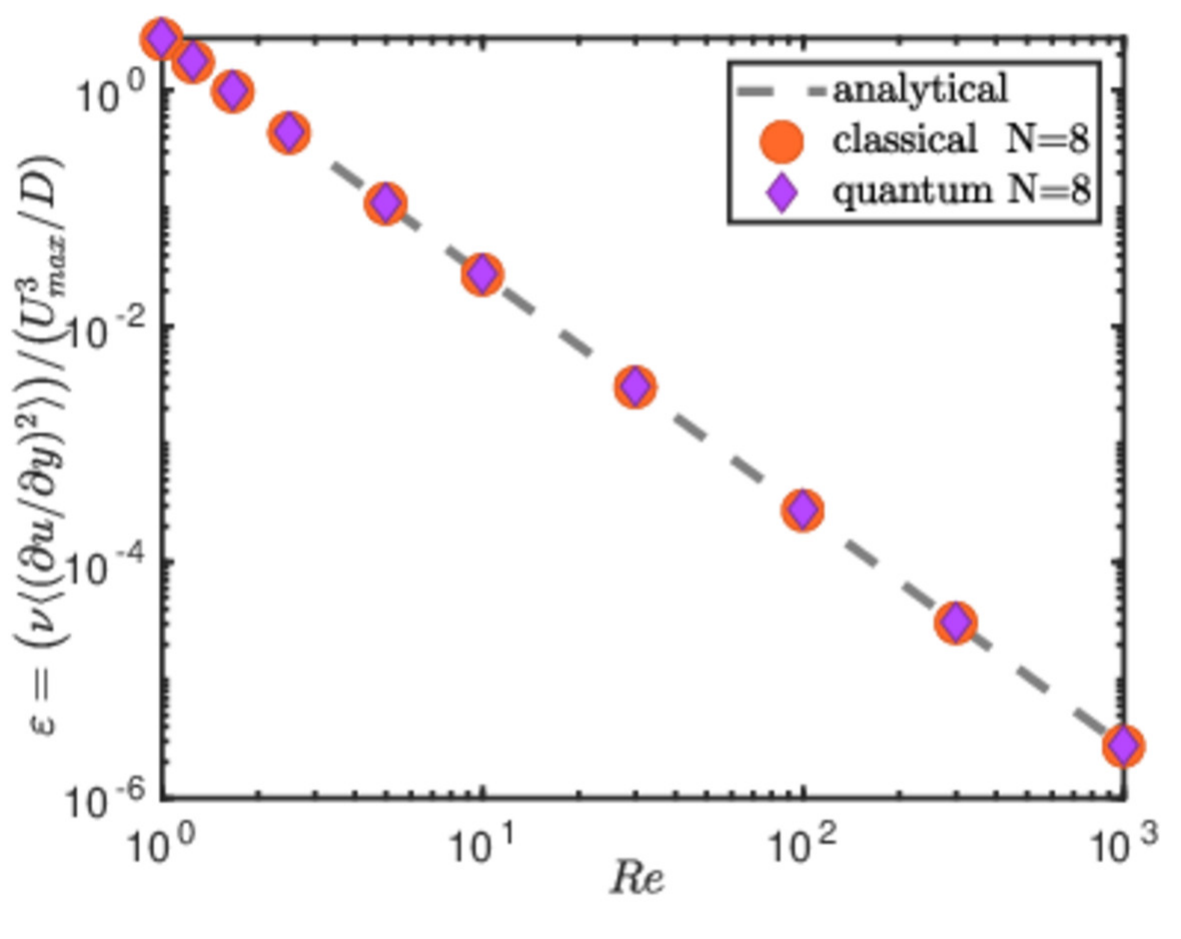}}
    
\caption{(a) \justifying shows the behavior of $\kappa$ with increasing $m$ and decreasing $\nu$. For $\nu < 0.01$, $\kappa$ saturates to a constant that is decided largely by $m$. This is expected since for Poiseuille flow, by appropriately scaling time with $\nu$ (using $t^{*} = \nu t/L^{2}$), $\nu$ can be absorbed into the equation. However, this is not the case when $\mathbf{C}\neq 0$ in eq. \ref{eq:governing1}, in which case $\nu$ would have a much stronger effect on $\kappa$.  (b) Using QFlowS, a small sample space of low dimensional matrices for varying $\kappa$ is solved to compute $T^{*}_{0}$ in each case, as shown in figure \ref{fig:error}(c), which yields a power law scaling of $T^{*}_{0}$ against $\kappa$, as marked in the figure. This feature can be used to predict $T^{*}_{0}$ for higher dimensional matrices. $T^{*}_{0}$ is insensitive to $\kappa$ below a threshold value of $\nu$ for these problems. Thin dashed lines indicate 99\% levels. (c) shows the mean viscous dissipation rate $\varepsilon$, normalized by the center-line velocity $U_{max}=(-\partial p/\partial x)/(4\nu)$), with respect to increasing $Re$ (decreasing $\nu$) computed from the steady-state analytical, classical and quantum solutions.}
\label{fig:scaling}
\end{figure*}
This favorable possibility arises because a subspace of the solution set could have had near exact representation using the given number of qubits, which lowers the overall L2 error $\leq 2^{-Q_{PE}}$. This means the minimum number of qubits needed to attain an error $\epsilon$ grows as a power law $(Q_{PE})_{min}\sim 2.92\epsilon^{-0.1158}$, as shown in the inset of figure \ref{fig:error}(d). If not for the right choice of $T^{*}_{0}$, for $Q_{PE}>3$, one would spend {a larger number of} qubits $\mathcal{O}(1.44\log(\epsilon^{-1}))$ to lower the overall error. We note that $\epsilon$ as computed here suppresses the error from finite differences, which is $\epsilon_{fd} \sim \mathcal{O}((\Delta y)^{2},\Delta t)$ and it's important to note that this error plagues both quantum and classical solutions.

Thus, being able to estimate $T^{*}_{0}$ fairly accurately reduces the overall computational resources required as well as the error, making it amenable for NISQ devices. Though there have been several analytical, asymptotic prescriptions for the choice of $T^{*}_{0}$ \cite{harrow2009quantum,childs2021high,scherer2017concrete}, the exact choice remains elusive. To shed better light, QFlowS is equipped with QPE optimizer subroutine which, on the basis of the nature of the flow problem and the numerical method (finite differences) used, estimates $T^{*}_{0}$ by very minimal classical pre-processing. Since $\kappa$ decides the range of eigenvalues and the invertibility of the matrix, it forms a common link that characterizes matrices for different systems with similar sparsity. Therefore, a relation that uniquely connects $\kappa$ with $T^{*}_{0}$ would make a reasonable basis for prescribing $T^{*}_{0}$ for different system configurations. We provide such a relation which, though generalizable in behavior, is specific to: (1) the class of elliptic and parabolic linear PDEs considered here; (2) finite difference based numerical formulations that give rise to either sparse, band diagonal, lower triangular, Toeplitz or circulant matrices. 

Since the one shot FE and BE seem to perform better than {iterative} BE, we take the matrix system of the former {(one shot FE)} for $N_{g}=10$ and characterize how $\kappa$ varies with the matrix size {(here we drop the constant factor $N_{g})$}, $m=\lceil\log_{2}(T/dt)\rceil$ and viscosity $\nu = 1/Re$. Figure \ref{fig:scaling}(a) shows for a specific case of $T=1$, $dt=0.001$, $\kappa$ grows as a stretched exponential with decreasing (increasing) $\nu$ ($Re$) and saturates for very low $\nu$, while for a fixed $\nu = 0.1$, $\kappa$ grows exponentially with $m$ as shown in the inset of figure \ref{fig:scaling}(a). However, the overall behavior of $\kappa$ with both $\nu$ and the system size $m$ is shown in figure \ref{fig:scaling}(b), where the $\kappa-\nu$ curves transition from exponential to stretched exponential fits as plotted for increasing $m$ obtaining the relation $\kappa \leq m(e^{-0.02m\nu}+2)$. This relation confirms that $\kappa$ is bounded and is not exponentially large. Here we explicitly highlight the dependence of $\kappa$ on $\nu$ and note that it is also bounded from above by $\kappa = 3(m+p+1)$, as given in \cite{liu2021efficient} for nonlinear PDEs that generally have higher $\kappa$ than linear PDEs. In both cases, $\kappa$ increases with $Re$ and $m$. 

We now proceed to compute $T^{*}_{0}$ for increasing $\kappa$, by extracting, as before, the $TQ$ phase diagram and obtaining the relation $T^{*}_{0} \sim -0.363\log(\kappa) +0.918$. We explore only a small zone of the phase diagram. The effect of $\nu$ is as expected, with $T^{*}_{0}$ decreasing and saturating for very low values. This relation between $T^{*}_{0}$ and $-\kappa$ obtained via simulations on QFlowS serves as a basis for choosing $T^{*}_{0}$ to perform accurate simulations for bigger circuit sizes as well. The earlier relation (despite a slight variation with matrix structure) forms a reasonable approximation for $T^{*}_{0}$ for the system considered here. {However, the process outlined here should be repeated to evaluate a problem-specific estimate of $T^{*}_{0}$; we have only presented a working level estimate of $T^{*}_{0}$ within a certain range to aid the eigenvalue estimation process. It} predicts optimal $T^{*}_{0}$ for Hamiltonian simulation algorithms for both QLSA-1 and QLSA-2. In case of QLSA-2, QFlowS also efficiently generates the set of the LCU coefficients for optimal performance. Further, even for other class of problems (different PDEs and discretization schemes), QFlowS's QPE optimizer could be employed to perform similar low cost classical pre-processing to suggest optimal $T^{*}_{0}$ for accurate and efficient fluid flow simulations. Further, barring minor quantitative differences, on performing a similar analysis on the Couette flow case we find that the qualitative outcome and inferences drawn are nearly the same as that of the Poiseuille flow case as seen here. The corresponding velocity profiles for the Couette flow are shown in \textit{SI Appendix} \ref{subsec:Couette}.

\section{Quantum Post-processing Protocol (QPP)}
\label{sec:QPP}
Once the velocity field is obtained, measuring it by repeated execution (excluding the requirements of quantum amplitude amplification \cite{brassard2002quantum}) of the quantum circuit ($\mathcal{O}(N)$ complexity) will compromise any quantum advantage and also introduce measurement errors. Here, we examine a QPP that produces just one real-valued output of the average viscous dissipation rate per unit volume, $\varepsilon=\nu\langle(\frac{\partial u}{\partial y})^{2}\rangle $. Given that we are equipped with an $U_{V}$ oracle \footnote{Oracles are black-boxes that represent a quantum algorithm or sub-routine with a specific operation. Here by $U_{V}$, we refer to QSP-1,-2 or QLSA-1,2).} that prepares a quantum register with the velocity field solution, we append to it a derivative module that computes first $\frac{\partial u}{\partial y}$ of the solution depicted in the circuit diagram shown in figure \ref{fig:Hybrid setup}(a). This is done by either (1) the LCU method where a finite difference matrix of first derivative is decomposed as linear combination of unitaries, or (2) a spectral method in which an IQFT is first applied to enter the conjugate space and the first derivative is now a simple scalar multiplication with the corresponding wavenumber, $k$. Finally the application of QFT transforms it back into real space. {Here, we implement the former method, given that the boundary conditions are non-periodic.}

At this point, if one is interested in general nonlinear functions such as trigonometric, logarithmic, square-root or higher powers, we implement the following procedure. The derivatives that are stored as quantum amplitudes are first converted into an $n_{m}$-bit binary representation using Quantum Analog-Digital Converter (QADC) \cite{mitarai2019quantum}. Following this step, a direct squaring algorithm outlined in \textit{SI Appendix} \ref{sec:QPP Append}, or a binary quantum arithmetic squaring circuit (an inverse of the square-root algorithm, which is more expensive than the former, see \cite{wang2020quantum}) is used to compute $(\frac{\partial u}{\partial y})^{2}$, which are finally converted back into amplitude encoding using Quantum Digital-Analog Converter (QDAC) \cite{mitarai2019quantum}. This algorithm requires $\mathcal{O}(1/\epsilon_{QPP})$ calls to the controlled-$U_{V}$ oracles that has a complexity of $\rm {\mathcal{O}~polylog(N/\epsilon)\kappa}$, thus an overall complexity of $\mathcal{O}(\text{polylog}(N/\epsilon)\kappa/\epsilon_{QPP} )$), one query to the bit-squaring algorithm with complexity $\mathcal{O}((\log_{2} N)^{2})$ and $\mathcal{O}((\log_{2} N)^{2}/\epsilon_{QPP})$ single- and two-qubit gates. {While computing the dissipation using QPP, we employ QFlowS with all its sub-modules, except the squaring step that is implemented as a separate circuit for proof-of-concept. In general, for larger system sizes each step of the QPP has to be implemented and simulated as distinct entities, since running them together exceeds the current simulator capacity.} The outline of the QPP algorithm introduced here along with circuit implementation is given in circuit \textit{SI Appendix} \ref{sec:QPP 
Append}.  

As a final step we apply a matrix $U_{avg}$ that computes the sum of derivatives at all points into one qubit, the measurement of which, along with a few ancillas, \footnote{Ancilla qubits are auxiliary or \textit{scratch-pad} qubits used to facilitate a specific quantum operation.} produces an output of the desired $\varepsilon$ (after some normalization and multiplication by $\nu$). Applying QPP on the quantum solution yields the variation of $\varepsilon$ with $Re$, computed near the steady-state, as shown in figure {\ref{fig:scaling}(c)}. {The transient from the uniform to the final, near-parabolic state is characterized by large gradients and warrants higher resolution. To keep the quantum resources within limits, we choose the well behaved, nearly steady-state profiles to compute dissipation. Further, even though in fig.~\ref{fig:scaling}(a) we explore large $Re \leq 10^{5}$ (where it is treated purely as a parameter to study bounds of $\kappa$), we compute dissipation only up to $Re=1000$. This is because, physically, (eq.\ref{eq:poiseuille flow}) is well suited to describe only low $Re$ flows. The normalized dissipation computed from the quantum solutions follows closely the classical and analytical results. The same finite difference based post-processing is used for all the three solutions to enable reasonable comparisons. In addition to the finite difference errors of classical solution, the quantum solution suffers from an additional error due to the quantum algorithms. Therefore, by trading-off with a smaller resolution (N=8), we obtain the quantum solutions by employing a larger number of QPE qubits, so that the overall error due to finite differences is at least comparable to its quantum counterparts (error in the quantum estimates, for instance with a modest $Q_{PE}=8$ and $Re \sim 10$, is $\epsilon \sim 10^{-3}$).} This could be made more accurate with more of the QPE qubits and the resolution $N_{g}$, as seen before. In essence, this demonstrates the possibility for computing quantities such as $\varepsilon$ effectively as a quantum post processing step.

\textit{SI Appendix} \ref{sec:QPP Append} includes a table of complexity formulae for the relevant QC subroutines used in this paper.

\section{Discussion}
\label{sec:Discussion}
We have demonstrated here a possible quantum algorithmic strategy and its full implementation using gate-based quantum circuits on QFlowS, to simulate Poiseuille and Couette flows in an end-to-end manner. First, we identify suitable quantum state preparation algorithms by considering the sparsity and functional forms of the initial velocity data being encoded. In CFD, it is generally admissible to choose a relatively simple form and a sparse initial condition, which would result in a relatively low cost of state preparation. QSP-1,2 could both be used as shown here, by assessing the form of input to encode initial and boundary conditions with sub-exponential complexity ($\mathcal{O}(\log(N_{be}))$ and $\mathcal{O}(kn)$, respectively), as well as to reinitialize instantaneous velocity fields. However, data that are dense with no functional form would force an exponential circuit depth. 

Second, using finite difference schemes, the governing equations were discretized to form linear system of equations, and solved by implementing QLSA-1 and -2, which are state-of-the-art, high precision algorithms with potentially up to an exponential advantage compared to classical schemes. Here, we have made a detailed analysis of the behavior of the velocity solutions and the attendant errors, which have revealed that FE outperforms BE. Further, we proposed the role of $T_{0}$ and discussed algorithms to prescribe the optimal value, $T^{*}_{0}$. The power-law and exponential form relations of ($\epsilon_{min}-Q_{PE}$) and ($T^{*}_{0}-\kappa$), respectively, given by QFlowS, forms a well-informed basis to choose $T^{*}_{0}$ and the minimum required qubits to perform accurate (up to $\epsilon_{min}$) and qubit-efficient flow simulations. Though QLSA-2 {avoids a standard} QPE and can provide exponential advantage in precision $\mathcal{O}(\text{polylog}(N/\epsilon)\kappa )$, other methods based on adiabatic QC \cite{subacsi2019quantum} could have potentially simpler implementations, while offering a similar performance, motivating further investigations.

To keep the discussion compact, the data reported here are taken mainly from QLSA-1; a similarly detailed discussion on QLSA-2 forms the bulk of the upcoming work. In QLSA-2 the critical factors computed are the coefficients for LCU, GPE parameters and a comparison between the Fourier and Chebyshev approaches.  Further, we have introduced a QPP protocol, where we propose the computation of the viscous dissipation rate $\varepsilon$, using a specific combination of QFT, IQFT, QADC, QDAC and bit-arithmetic, with an overall complexity that scales as $\mathcal{O}(\text{polylog}(N/\epsilon)\kappa/\epsilon_{QPP} + ((\log_{2} N)^{2}/\epsilon_{QPP}) )$. The QPP introduces an extra $\mathcal{O}(1/\epsilon_{QPP})$ scaling that brings down the performance of QLSA-2 to the level of QLSA-1 (if $\epsilon \approx \epsilon_{QPP}$). Added to this, for purposes of quantum amplitude amplification, QFlowS is capable of repeating circuit runs in parallel (currently tested  up to $\sim$8000 shots, {where a shot simply means a repeated run of the quantum circuit to extract quantum state statistics}).

We should point out that this method avoids measuring the entire velocity field, thus protecting it from compromising the quantum advantage and escalating possible measurement errors. We observe that $\varepsilon$ computed from the resulting quantum simulations captures the known analytical results. This method can be extended to compute other nonlinear functions of the velocity field.

In summary, {we have demonstrated} a complete implementation of an end-to-end algorithm to perform fluid flow simulations using QC, which paves the way for future QCFD simulations of both linear and nonlinear flows. We have also introduced here a new quantum simulator package QFlowS---designed mainly for CFD applications. It is important to note that we have not addressed other key challenges such as noise and quantum error correction, which we emphasize are critical to simulations on near term quantum devices. The complexities of the algorithms provided here (see \textit{SI Appendix} \ref{sec:QPP Append}) are estimates. {Ongoing efforts include investigations of higher order finite difference schemes, comparative studies of different amplitude amplification approaches, detailed error} and complexity analyses, along with computing exact gate counts and circuit depths. 
Extending these methods and tools to nonlinear systems such as Burgers equations and Navier-Stokes equations also part of this effort.

\subsection*{Data availability}\label{subsec:data avail} 
Data, Materials, and Software Availability. The package QFlowS will be made is available as an open-source package on GitHub \cite{qflowsgit}. Readers interested in using QFlowS can write to the authors. All other data are included in the manuscript and/or SI Appendix.

\acknowledgements{We wish to thank Dhawal Buaria (NYU), Yigit Subaśi (LANL), Jörg Schumacher (TU Ilmenau), Balu Nadiga (LANL), Patrick Rebentrost (CQT), Stefan Wörner (IBM) and Philipp Pfeffer (TU Ilmenau) for insightful discussions, and the referees for helpful comments. S.S.B acknowledges the computational resources provided by the NYU Greene supercomputing facility on which these simulations were performed.}

\appendix
\section{Numerical method}
\label{sec:Numerical method}
\subsection{Finite difference} 
\label{subsec:FDM}
\subsubsection{Spatial discretization} The well-known 2\textsuperscript{nd} order central difference scheme is used to discretize the flow domain into $N_{g}$ equidistant grid points, as shown in figure S1(a) 
($u = [u_{1},u_{2}, \cdots ,u_{N_{g}}]$), with grid spacing $\Delta y=h=1/(N_{g}+1)$, admitting a discretization error $\sim \mathcal{O}(\Delta y^{2})$. Since the velocity is known at the boundaries, one solves only for the $N_{g}-2$ unknown internal grid points. Thus the Laplacian operator can be written as

\begin{equation}
    \Delta u = \frac{u(y_{i}+h)- 2u({y_{i}})+u(y_{i}-h) }{h^{2}} + \text{h.o.t} .
\label{eq:CDS}
\end{equation}

Now denoting this discretization operator as matrix $\mathbf{A}$ and letting the pressure gradient form a constant vector $\mathbf{f}$, we can rewrite eq. 2 (in the main text) as

\begin{equation}
    {\frac{du}{dt}} = \mathbf{A}u + \mathbf{f} .
\end{equation}

\subsubsection{Temporal discretization:} To integrate in time, the temporal domain $t \in [0,T]$ is discretized into $m = T/\Delta t$ time steps using two different schemes both admitting an error  $\sim \mathcal{O}(\Delta t)$:
\begin{enumerate}
    \item Backward Euler or Implicit method: This discretizes the time derivative as
    \begin{equation}
        \frac{u^{j+1}-u^{j}}{\Delta t} = \mathbf{A}u^{j+1} + \mathbf{f},
    \end{equation}

and gives the matrix equation
\begin{equation}
    \mathbf{A}_{be1}\bar{u} = \mathbf{b}_{be1},
\end{equation} which needs to be inverted recursively to obtain the velocity field at every time step, where $\mathbf{A}_{be1} = -(\mathbf{A}\Delta t - \mathbf{I}) $, $\bar{u} = u^{j+1}$ and $\mathbf{b}_{be1} = u^{j} + \mathbf{f}\Delta t $. This scheme is known to be unconditionally stable with any choice of the size of $\Delta t$. \\

We also set up an alternative matrix equation 
\begin{equation}
    \mathbf{A}_{be2}\bar{u} = \mathbf{b}_{be2},
\end{equation} for all time steps as shown in eq. \ref{eq:BE_OBM}. $(A_{be2})_{ij} = -(\mathbf{I} + A\Delta t) ~ \forall ~ i=j $. However, $\forall~ i\leq m$, $ (A_{be2})_{ij} = -\mathbf{I} ~\text{for}~ j=i-1 $ and $\forall~ i\geq m$,  $ (A_{be2})_{ij} = -\mathbf{I} ~\text{for}~ j=i-1 $. Further, $(b_{be2})_{i} = \{u_{in} \forall ~ i=0 $; $= -\mathbf{f}\Delta t 
~\forall~ 0<i\leq m $; and $= 0~ \forall ~ i>m\}$.

\item Forward Euler or Explicit method: Here the discretization is given by

\begin{equation}
        \frac{u_{i}^{j+1}-u_{i}^{j}}{\Delta t} = \mathbf{A}u_{i}^{j} + \mathbf{f} ,
    \end{equation}

which leads to the matrix equation,

\begin{equation}
    \mathbf{A}_{fe}\tilde{u} = \mathbf{b}_{fe},
\end{equation}

where $\mathbf{A}_{fe}$, has a double-banded structure as written in eq. \ref{eq:FE_OBM}, $(A_{fe})_{ij} = \mathbf{I}~ \forall ~ i=j $ (see below). However, $\forall~ i\leq m$, $ (A_{fe})_{ij} = -(\mathbf{I} + A\Delta t) ~\text{for}~ j=i-1 $ and $\forall~ i\geq m$,  $ (A_{fe})_{ij} = -\mathbf{I} ~\text{for}~ j=i-1 $. Further, $(b_{fe})_{i} = \{u_{in} \forall ~ i=0 $; $= \mathbf{f}\Delta t 
~\forall~ 0<i\leq m $; and $= 0~ \forall ~ i>m\}$.\\
\end{enumerate}
Equations \ref{eq:BE_OBM} and \ref{eq:FE_OBM} (see below) thus unroll all the time steps into one big matrix of dimensions $(m+p+1)N_{g} \times (m+p+1)N_{g}$, 
thus solving for the velocity $\tilde{u} = [u^{0},u^{1},\cdots,u^{m+p}]$ at all times in one shot, where every $u^{j}$ is the full field at all grid points. The total time $T$ is discretized into $m+p$ time steps, where one chooses a large enough $p$, such that every $u^{j}_{i} = u^{j+1}_{i}$ for $j \in [m+1,m+p]$, which implies that, after the attainment of a steady state, the solution produces $p$ copies of the the steady state solution. This is done only to improve the measurement probability of the post-selected state \cite{liu2021efficient} and does not affect the solution itself. This method is stable only for an appropriate choice of Courant number (von Neumann stability criteria), $\alpha = \frac{\Delta t}{(\Delta y)^{2}} < 0.5$, making it conditionally stable and specific to the PDE under discussion, which therefore also decides the upper bound on the largest admissible $\Delta t$. 
\begin{figure*}[htb!]
       \centering
        \subfloat[]{
            \includegraphics[scale=0.37]{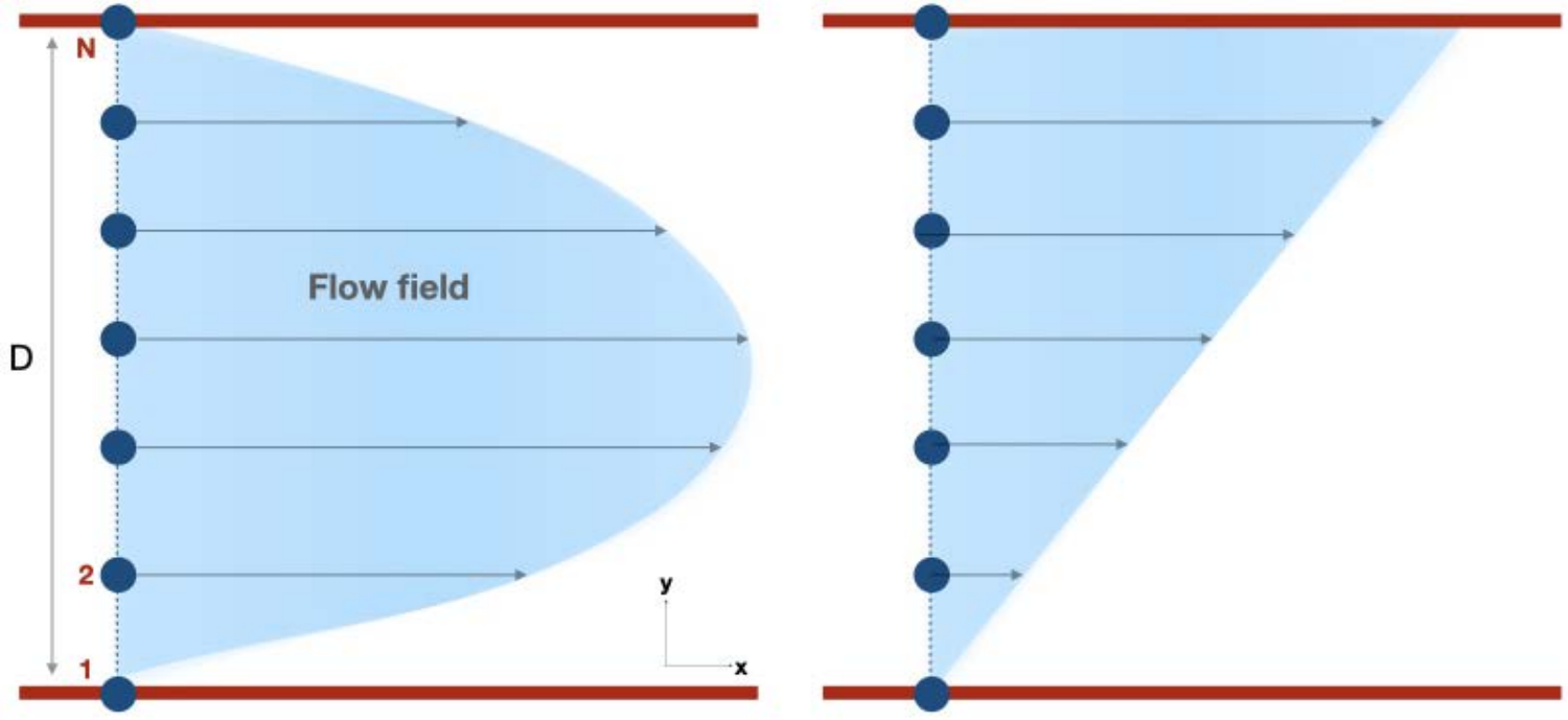}
       }
        \subfloat[]{
          \includegraphics[scale=0.43]{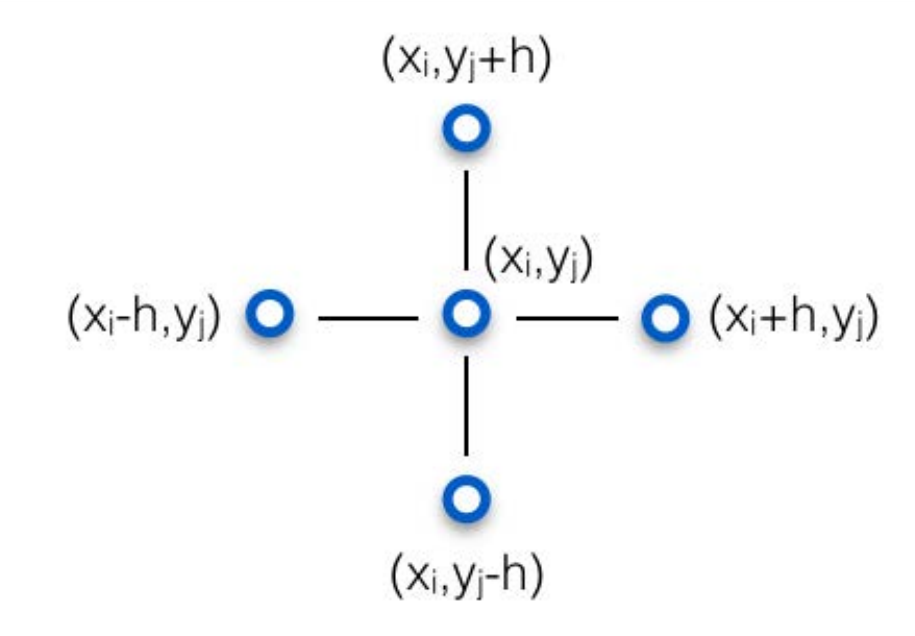}
        }
       \caption{\textbf{(a)} \justifying Shows the computational flow field setup: The flow is confined between two no-slip boundaries separated by a distance $D = 1$, with $x$ and $y$ as the streamwise and wall-normal directions, respectively. The left and right panels in (a) show the schematics of steady state velocity profiles of the Poiseuille and Couette flows, respectively. The finite difference scheme discretizes the flow domain along $y$ into $N$ grid points with separation $\Delta y$, at which the velocities are computed. \textbf{(b) } shows the 2\textsuperscript{nd} order central difference stencil in 2D. For the 1D case, only the points in the y-direction are considered and the resulting approximation to the Laplacian operator estimated at a given point $y_{i}$ using the two points on either side of it is given in eq. \ref{eq:CDS}. }
       \label{fig:flow setup}
    \end{figure*}
    \begin{figure*}[htpb!]
\begin{equation}
\begin{bmatrix}
\mathbf{I} & 0 & \cdots &  &  &  &\\
-\mathbf{I} & (\mathbf{I} - \mathbf{A}\Delta t) &  & &  & 
\\
0 & \ddots & \ddots    &\\
\vdots &  & -\mathbf{I}  & (\mathbf{I} - \mathbf{A}\Delta t) & &\\
 & & &-\mathbf{I} & \mathbf{I} & &\\
 &  &  &   & \ddots & \ddots\\
 &  &  & & & -\mathbf{I} & \mathbf{I} \\

\end{bmatrix} \begin{bmatrix}
u^{0}\\u^{1}\\ \vdots\\ u^{m}\\ u^{m+1}\\ \vdots\\ u^{m+p}\\ 
\end{bmatrix} = \begin{bmatrix}
u^{in}\\b_{0}\\ \vdots\\ b_{m-1}\\ 0\\ \vdots\\ 0\\ 
\end{bmatrix} 
\label{eq:BE_OBM}
\end{equation}
\end{figure*}
\begin{figure*}[htpb!]
\begin{equation}
\begin{bmatrix}
\mathbf{I} & 0 & \cdots &  &  &  &\\
-(\mathbf{I} + \mathbf{A}\Delta t) & \mathbf{I} &  & &  & 
\\
0 & \ddots & \ddots    &\\
\vdots &  & -(\mathbf{I} + \mathbf{A}\Delta t) & \mathbf{I} & &\\
 & & &-\mathbf{I} & \mathbf{I} & &\\
 &  &  &   & \ddots & \ddots\\
 &  &  & & & -\mathbf{I} & \mathbf{I} \\

\end{bmatrix} \begin{bmatrix}
u^{0}\\u^{1}\\ \vdots\\ u^{m}\\ u^{m+1}\\ \vdots\\ u^{m+p}\\ 
\end{bmatrix} = \begin{bmatrix}
u^{in}\\b_{0}\\ \vdots\\ b_{m-1}\\ 0\\ \vdots\\ 0\\ 
\end{bmatrix} 
\label{eq:FE_OBM}
\end{equation}
\end{figure*}

\subsection{Poiseuille flow}
\label{subsec:Poiseuille}
The discussion of the Poiseuille flow is provided in the main text and will not be repeated here. 

\subsection{Couette flow}
\label{subsec:Couette}
\begin{figure*}[htb!]

\subfloat[]{
    \includegraphics[trim={0.9cm 0.2cm 1.5cm 0.9cm},clip=true,scale=0.35]{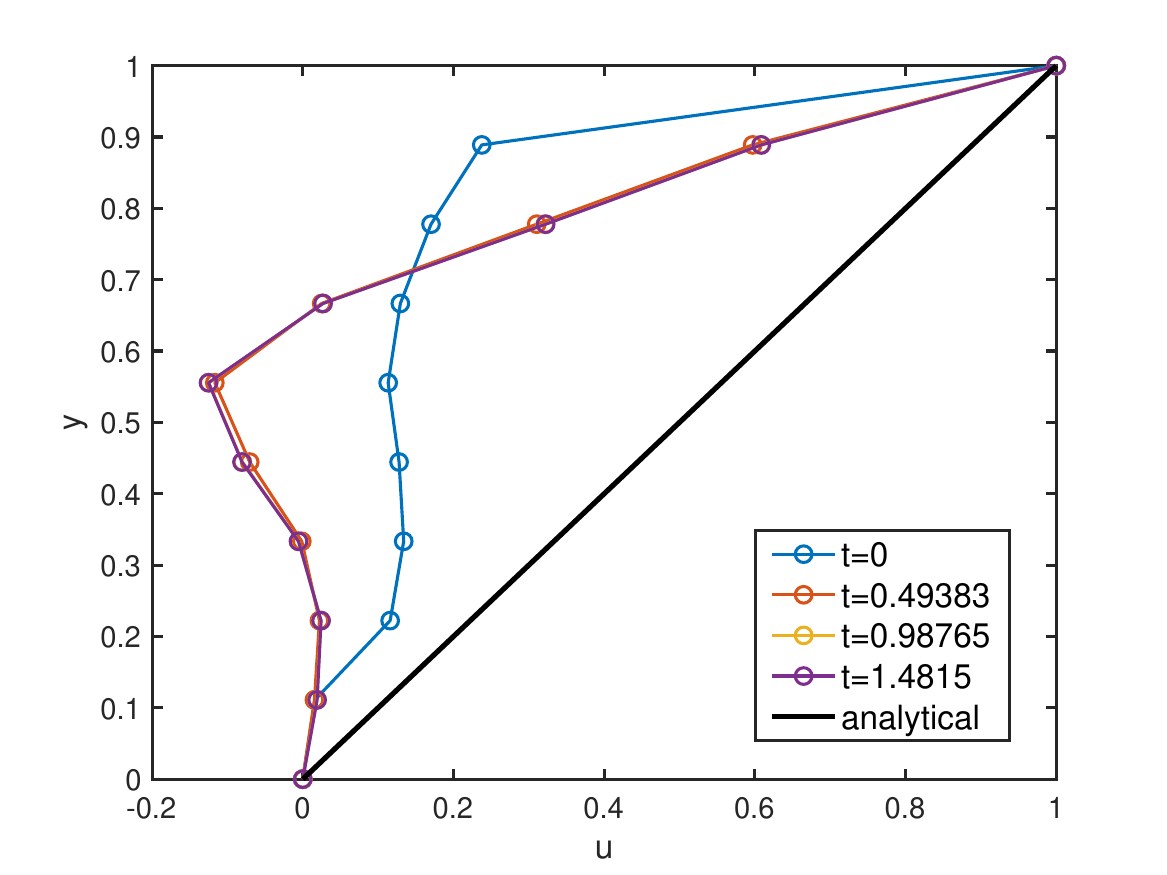}  }
    \subfloat[]{
         \includegraphics[trim={0.9cm 0.2cm 1.5cm 0.9cm},clip=true,scale=0.35]{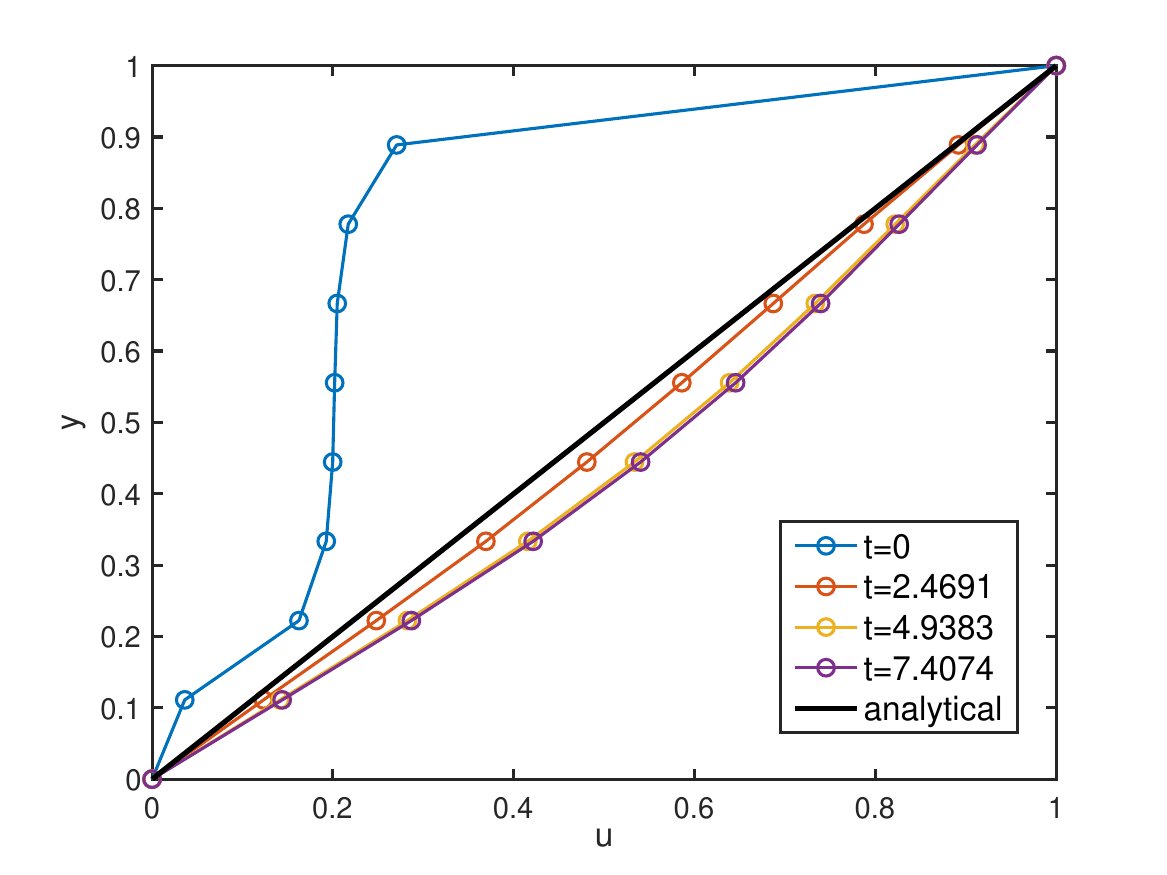}
    }
    \subfloat[]{
         \includegraphics[trim={0.9cm 0.2cm 1.5cm 0.9cm},clip=true,scale=0.35]{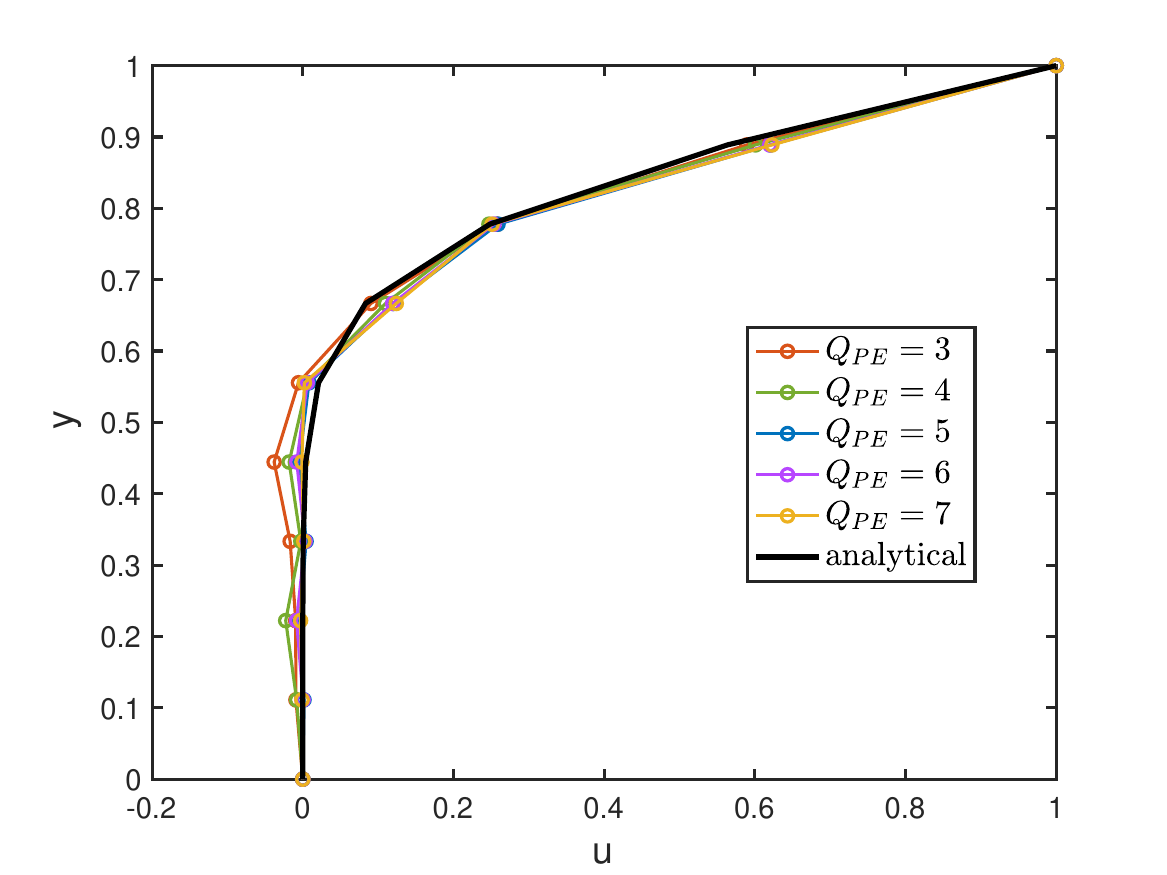}
    }
\centering{  
\subfloat[]{
         \includegraphics[trim={0.9cm 0.2cm 1.5cm 0.9cm},clip=true,scale=0.35]{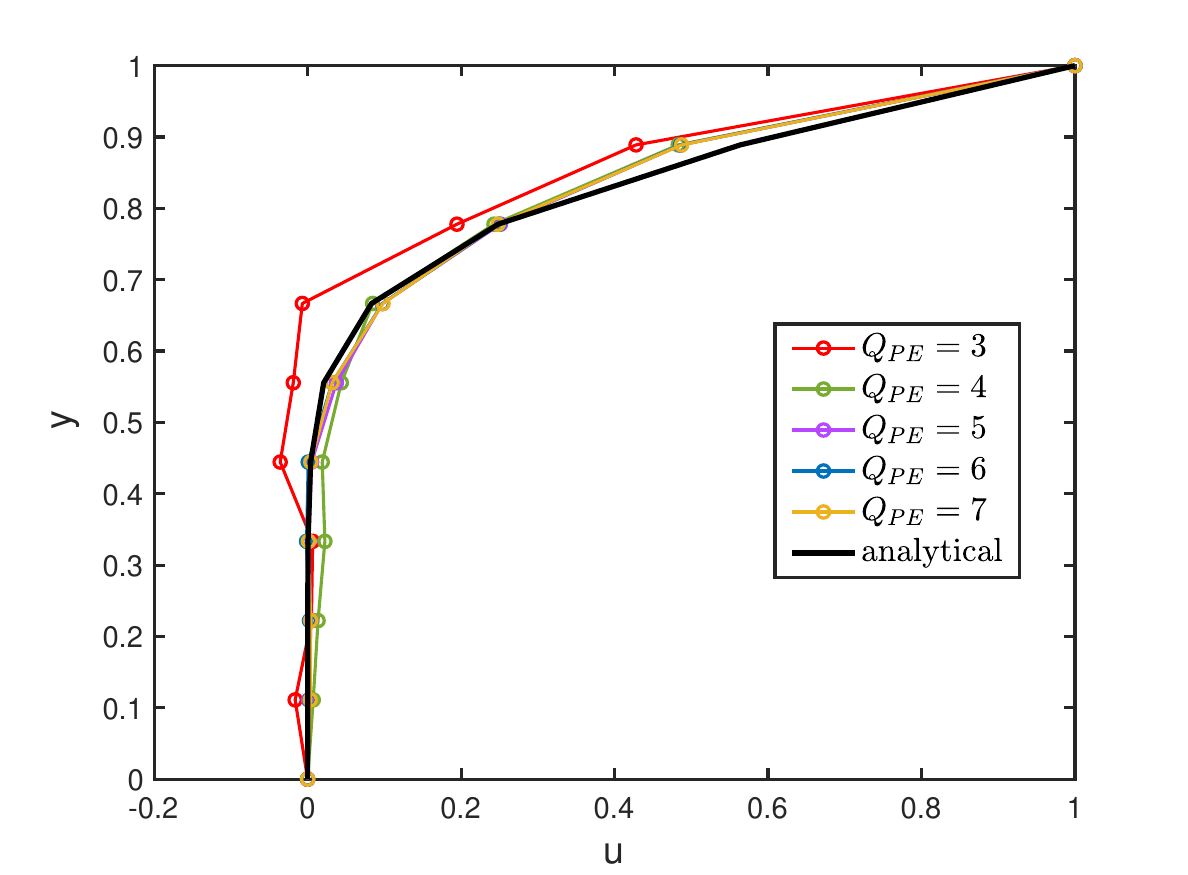}
    }
    \subfloat[]{
         \includegraphics[trim={0.5cm 0 1.3cm 0.7cm},clip=true,scale=0.35]{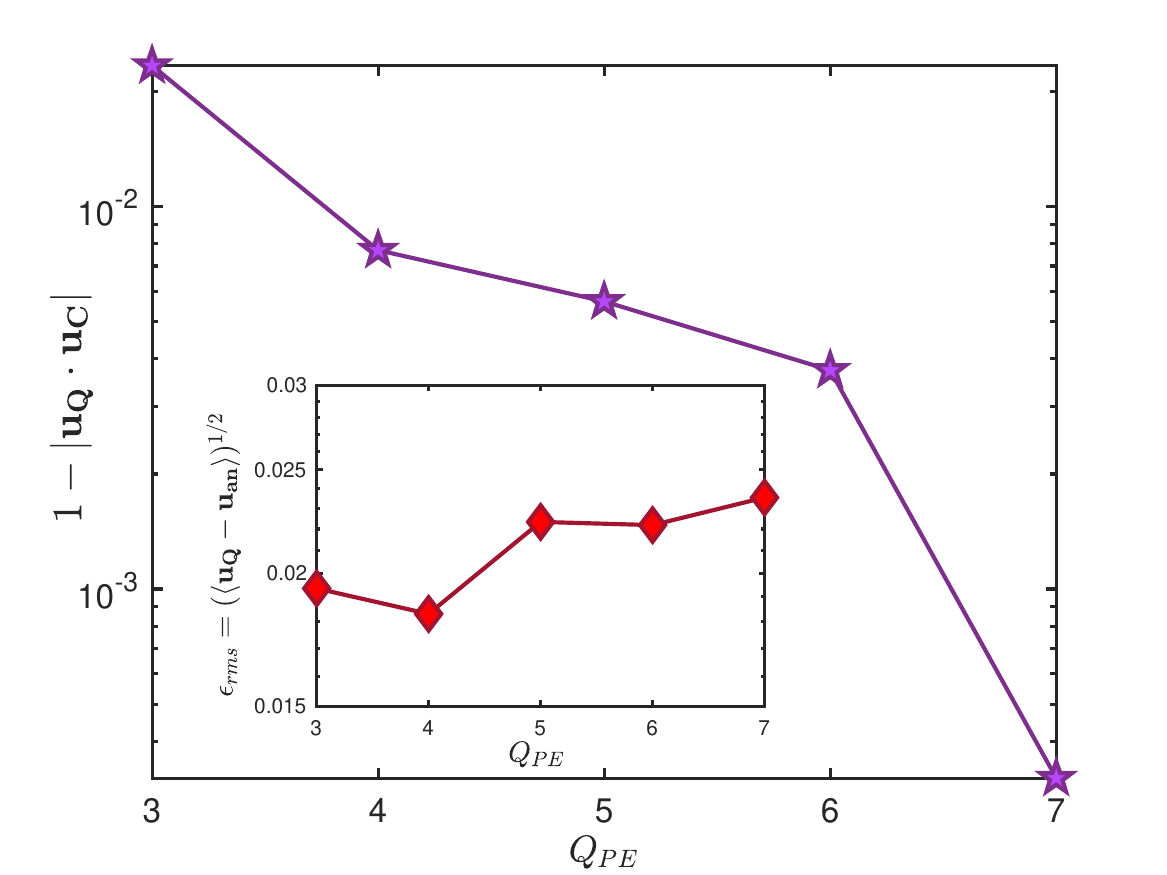}
    }
    }
\caption{\justifying (a) and (b) shows the quantum simulation of the flow field evolving forward in time towards steady state (analytical solution shown as solid black line) using BE scheme that use 7 and 12 qubits (3 and 8 QPE qubits ($Q_{PE}$), respectively), with $\mathbf{u(y,0)})=0$, $N=10$, $Re = 10$, $\partial p /\partial x = 0$ and $dt = 0.01$. The accuracy of the converged solution improves for higher $Q_{PE}$ similar to the Poiseuille flow case. Here the velocity field is solved for at every time step; (c) shows increasingly accurate converged solutions with increasing $Q_{PE}$, but solved using the FE scheme where the velocity field is solved for all time steps in one shot, and only the final solution is extracted. Here $\alpha = 0.5$ is set to meet the the von Neumann stability criterion and the parameter $T^{*}_{0} = 5.0$ is fixed. It is important to note that, the same $T^{*}_{0}$ works for both Poiseuille flow and Couette flow independent of the boundary conditions, since it depends only on $\kappa$ and form of the matrix. (d) shows the one-shot method, but it is solved with a BE scheme for $T^{*}_{0} = 6.5$.
(e) shows both (1-fidelity) and $\epsilon_{rms}$ (inset) with increasing $Q_{PE}$ from 3 to 7 computed for the one-shot FE case.} 
    \label{fig:velocity field 2}
\end{figure*}

Following the same procedure as for Poiseuille flow, the Couette flow ($\mathbf{u}(0,t)=0, \mathbf{u}(1,t)=1$) can be captured accurately as seen in the flow profiles of figure \ref{fig:velocity field 2}(a-d). Similar inferences to those discussed in the main text are applicable for this case as well. We wish to highlight two possible measures of accuracy: (i) the fidelity = $|\mathbf{u_{Q}\cdot u_{C}}|$ and (ii) RMS error (with respect to analytical solution), given by $\epsilon_{rms}=(\langle \mathbf{u_{Q}-u_{an} }\rangle)^{1/2}$, both of which are plotted in figure \ref{fig:velocity field 2}(e) for a one shot FE scheme with a nearly accurate estimate of $T^{*}_{0}$. We can clearly observe that with increasing $Q_{PE}$, the fidelity increases, though the $\epsilon_{rms}$ has a weakly increasing trend. This indicates that higher fidelity does not necessarily indicate better $\epsilon_{rms}$, hence perhaps not be the most robust measure of performance when solving physical, fluid mechanical problems. Higher fidelity only indicates larger overlap of the quantum solutions with respect to the classical inversion solution, which itself is technically erroneous due to finite discretization and truncation errors. Even if fidelity=1, the error is still bounded by $\sim\mathcal{O}((\Delta y)^{2},\Delta t)$. The solution for the one shot case is a wavefunction that encodes all time steps, but we extract only the final time step. The fidelity is an overall measure of this solution but does not quantify the accuracy of the vector subspace in the final step. Also, the fidelity in the iterative BE case drops for every time step, since a dynamical $T^{*}_{0}$ is not chosen for every time step. Thus higher fidelity can be seen only as an indicator of whether the chosen quantum parameters for QLSA are at least in a reasonable ballpark.

\section{QFlowS - A brief overview of the package}
\label{sec:QFlowS overview}
\begin{figure*}[htpb!]
        \centering
    \includegraphics[trim={0cm 4cm 0cm 6.5cm},clip=true,width=0.9\textwidth]{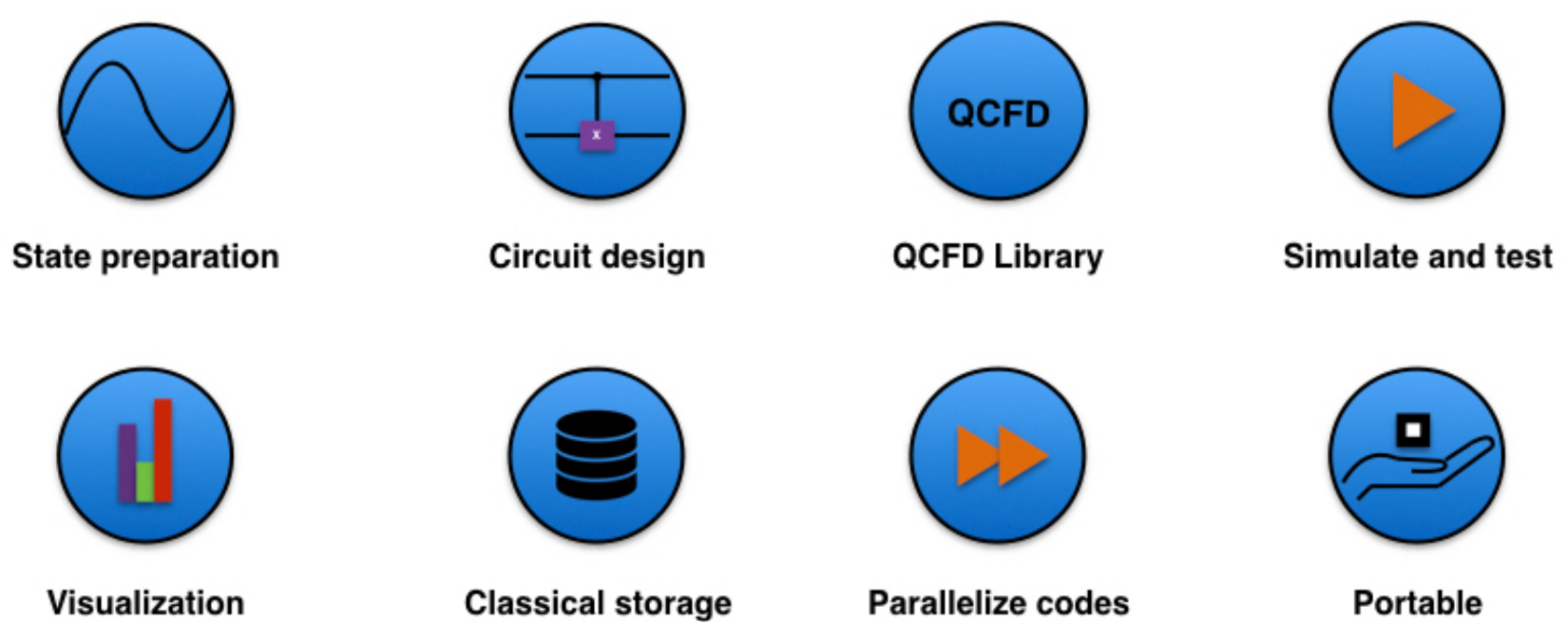}
        \caption{Overview of the QFlowS features}
\end{figure*}
QFlowS is a specialized high performance quantum simulator that enables setting up CFD problems in the QC format seamlessly. We summarize here briefly the various features of QFlowS as schematically depicted in figure \ref{fig:flow setup}(a). 

\begin{enumerate}
    \item Qubits and quantum states: Qubits which are quantum analogues of classical bits, form the fundamental units of information storage and are represented by quantum states that follow rules of quantum mechanics. Mathematically, they form elements of a complex vector space (Hilbert space $\mathbb{H}$ $(\in \mathbb{C}^{n})$). An n-qubit state of the quantum computer, formed by taking tensor products of single qubit states $|\psi\rangle$, is given by
    \begin{equation}
        |\psi\rangle^{\otimes n} = \sum^{2^n}_{i=1}c_{i}|u_{i}\rangle, ~~ c_{i} \in \mathbb{C},
    \end{equation} which encodes $2^{n}$ complex values $c_{i}$, in the basis $|u_{i}\rangle$, that are stored as 1-dimensional arrays on QFlowS. The memory required should ideally scale linearly with wavefunction size ($\approx 16\times2^{n}$ bytes with double precision), but there is an overhead due to the need to store quantum circuit instructions. Currently, QFlowS offers simulation capabilities with up to 30+ qubits, which span vector spaces with dimensions of the order $\sim 10^{9}$. For performing parallelized simulations, these quantum states are either (a) loaded on a large memory single node architecture (with or without GPU) dealt with an OpenMP style parallelization, or (b) distributed onto different processors for an MPI style execution. Both methods were tested initially but the former method was favored because of the simplicity of implementation and lower communication overheads. Some key operations on quantum states that can be done with QFlowS include:
    \begin{enumerate}
        \item Quantum State Preparation (QSP) - To initialize any arbitrary states or states with special features. This is detailed further in \textit{SI Appendix} \ref{sec:QSP Append};
        \item Quantum state tomography and amplitude estimation - To estimate the amplitude of the final quantum state by reconstructing the state using different tomography techniques \cite{nielsen2002quantum};
        \item Quantum state characteristics - To compute other useful properties of the state such as density matrices, entanglement and norm. 
    \end{enumerate}
    
    \item Quantum gates and circuits: Quantum gates are given by unitary operators $U$ ($UU^{\dag} = U^{\dag}U = \mathcal{I}$), that are essentially rotation matrices, which collectively form a quantum circuit that manipulates quantum information in a specific way. The quantum circuit can be viewed as a tensor product of all the single qubit gates that forms a matrix of the size $2^{n}\times2^{n}$ for an n-qubit circuit. On QFlowS, the quantum gates are not implemented as matrix operations, but as algebraic operations. The exact transformation caused by different gates is translated into vector operations that affects the specific coefficients, but done in parallel on multiple cores. This makes the simulator more efficient with respect to both memory and speed. For example, a Hadamard gate and a NOT gate acting on a two qubit state shown in figure S4 is given by
    \begin{figure*}[htb!]
        \begin{equation}
    \vcenter{\hbox{\begin{minipage}{5cm}\includegraphics[width=1.2\textwidth]{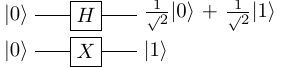}\captionof{figure}{2 qubit circuit}\end{minipage}}}
\qquad\qquad\hspace{0.1cm}
U = H_{1}\otimes X_{0} = \frac{1}{\sqrt{2}}\begin{bmatrix}
0 & 1 & 0 & 1\\
1 & 0 & 1 & 0\\
0 & 1 & 0 & -1\\
1 & 0 & -1 & 0\\
\end{bmatrix}\implies \begin{bmatrix}
u_{0}\\
u_{1}\\
u_{2}\\
u_{3}\\
\end{bmatrix} \mapsto \frac{1}{\sqrt{2}}\begin{bmatrix}
u_{1}+u_{3}\\
u_{0}+u_{2}\\
u_{1}-u_{3}\\
u_{0}-u_{2}\\
\end{bmatrix}
\end{equation}

    \end{figure*}

The action of such a circuit brings about a transformation on the state, as given by eq.~11. Such a transformation could be algebraically dealt with as follows. In an n-qubit circuit, a Hadamard gate acting on the qubit $q$ is $H_{q}$, for which $u_{i} \mapsto \frac{1}{\sqrt{2}}(u_{i}\pm u_{i+2^{n-q-1}})$. This operation is performed on vector elements by the parallel processing of causally disconnected (unentangled) gates and vector sub-spaces without having to store or multiply the large $2^{n}\times 2^{n}$ matrices. QFlowS can successfully handle circuits with $\sim 10$ million multi-controlled and two-level gates even on a standard workstation with 10 core CPU, 1TB memory and 32GB RAM.
    \item Algorithm library and portability: QFlowS is docked with several standard quantum subroutines or algorithms such as Quantum Fourier Transform (QFT) and Quantum Phase Estimation (QPE) that can be readily used in any new circuit. Along with that, QFlowS has classical CFD tools such finite difference (FDM) , finite volume (FVM) and boundary element methods (BEM), implicit and explicit time stepping methods, predictor-corrector methods and linearization methods such as Homotopy Analysis Method (HAM) and Carleman method---to highlight a few. These classical subroutines generate appropriate matrices and vectors in formats that can be imported seamlessly by the quantum algorithms. This package will offer easy portability onto both local workstations and supercomputers.
    
    \item Visualization: With QFlowS, the quantum states and quantum circuits can also be visualized with an in-built state histogram builder and circuit drawer, with which algorithms and states can be visualized and verified for correctness to simulate and test the circuits.

\end{enumerate}

\section{Quantum State Preparation}
\label{sec:QSP Append}
QFlowS is equipped with a Quantum State Preparation (QSP) library that includes algorithms that correspond to certain classes of states with specific properties. We briefly elucidate two such algorithms used here. When the state that is being initialized on a quantum computer has a specific functional form such as log-concavity, it is known that it can be prepared efficiently \cite{grover2002creating,prakash2014quantum,vazquez2022enhancing,vazquez2021efficient}. The original algorithm proposed in \cite{grover2002creating} proceeds as follows. Consider an arbitrary n-qubit initial state given by
\begin{equation}
    |\psi\rangle = \sum_{i=0}^{2^{n}-1}\sqrt{p_{i}}|i\rangle,
\end{equation}
where $p_{i}$ is the $i$-th region of discretely sampled elements/regions from a log-concave probability distribution function $p(\omega)$. With the existence of an efficient classical subroutine to perform partial sums given by
\begin{equation}
    p_{i} = \int^{\omega^{i}_{R}}_{\omega^{i}_{L}} p_{\omega}d\omega ,~~~~~~
    p_{iL} = \int^{(\omega^{i}_{R}-\omega^{i}_{L})/2}_{\omega^{i}_{L}} p_{\omega}d\omega,
    \label{eq:region probablities}
\end{equation} where $p_{i}$ and $p_{iL}$ are probabilities of the point lying in the entire region $i$ and the left half of the region $i$, respectively, we can construct a circuit that prepares an n-qubit state for $k<m$ as
\begin{equation}
 |\psi^{(k)}\rangle = \sum_{i=0}^{2^{k}-1}\sqrt{p^{(k)}_{i}}|i\rangle.
\end{equation}
Now we can further discretize this to yield the state
\begin{equation}
 |\psi^{(k+1)}\rangle = \sum_{i=0}^{2^{k+1}-1}\sqrt{p^{(k+1)}_{i}}|i\rangle
\end{equation}
by the following steps. Given the ability to compute the quantities in eq.~\ref{eq:region probablities}, we can compute the conditional probability function $f_{k}(i)$
\begin{equation}
    f_{k}(i) = \frac{p_{iL}}{p_{i}}.
\end{equation} With this we now compute the next level of discretization by constructing a quantum arithmetic circuit that performs
\begin{equation}
    |i\rangle|0\rangle \mapsto |i\rangle|\theta_{i}\rangle,
\end{equation} where $\theta_{i} =  \arccos(\sqrt{f_{k}(i)})$. Further, by adding an ancillary qubit ($(k+1)$-th qubit) we perform controlled $R_{y}(\theta_{i})$ rotation gates (controlled on $\theta$); uncomputing the second register we get

\begin{align}
    \sqrt{p^{(k+1)}_{i}}|i\rangle|\theta_{i}\rangle|0\rangle &\mapsto \sqrt{p^{(k)}_{i}}|i\rangle|\theta_{i}\rangle(\cos \theta_{i}|0\rangle+\sin \theta_{i}|1\rangle) 
      \nonumber \\ &\equiv \sum_{i=0}^{2^{k+1}-1}\sqrt{p^{(k+1)}_{i}}|i\rangle = |\psi^{(k+1)}\rangle.
\end{align}

\begin{figure*}[htb!]
        \centering
        \subfloat[]{
    \includegraphics[trim={0cm 0cm 0cm 0cm},clip=true,width=1\textwidth]{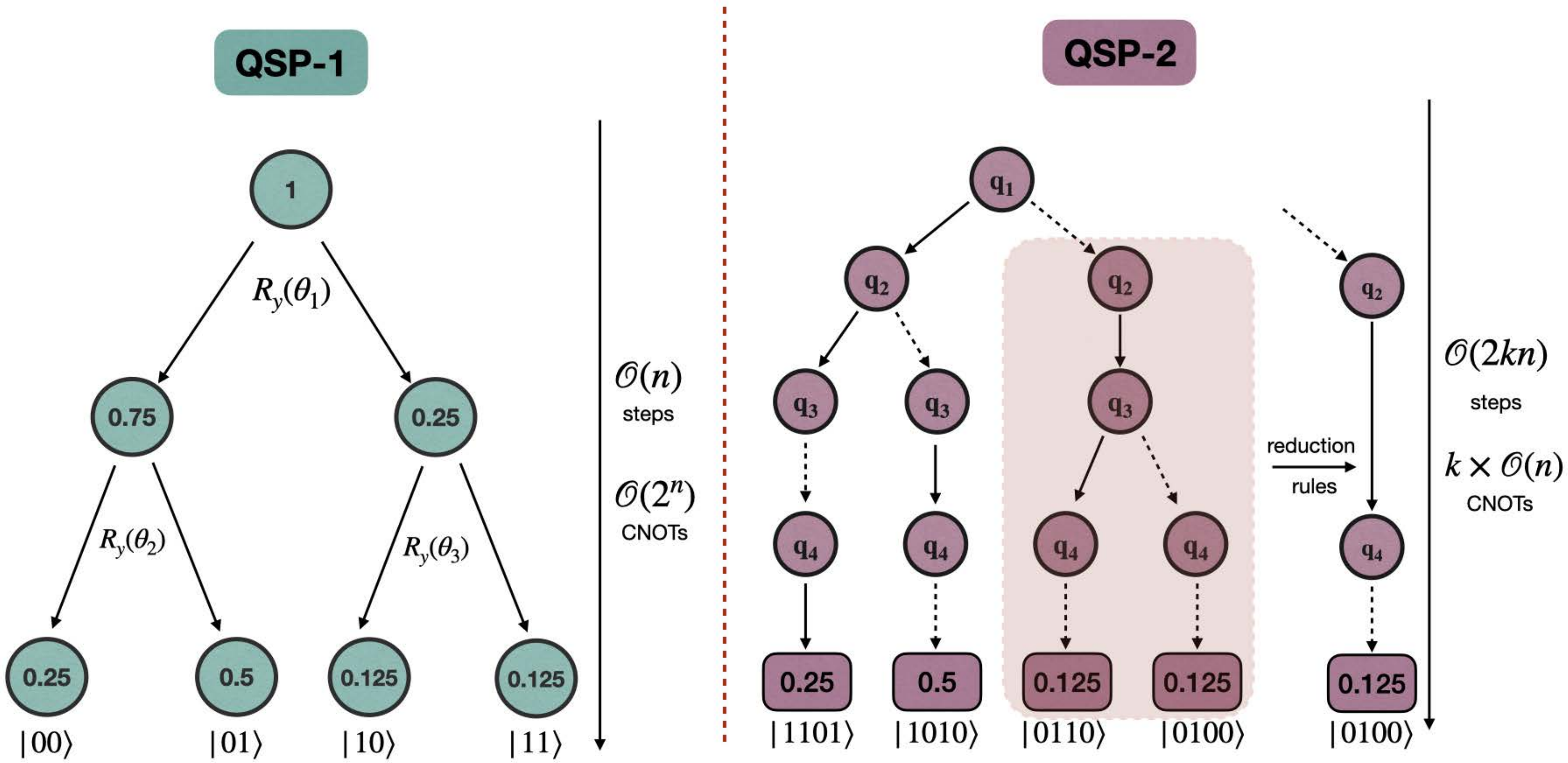}}\\
    \subfloat[]{\includegraphics[trim={0cm 7cm 0cm 7cm},clip=true,width=0.8\textwidth]{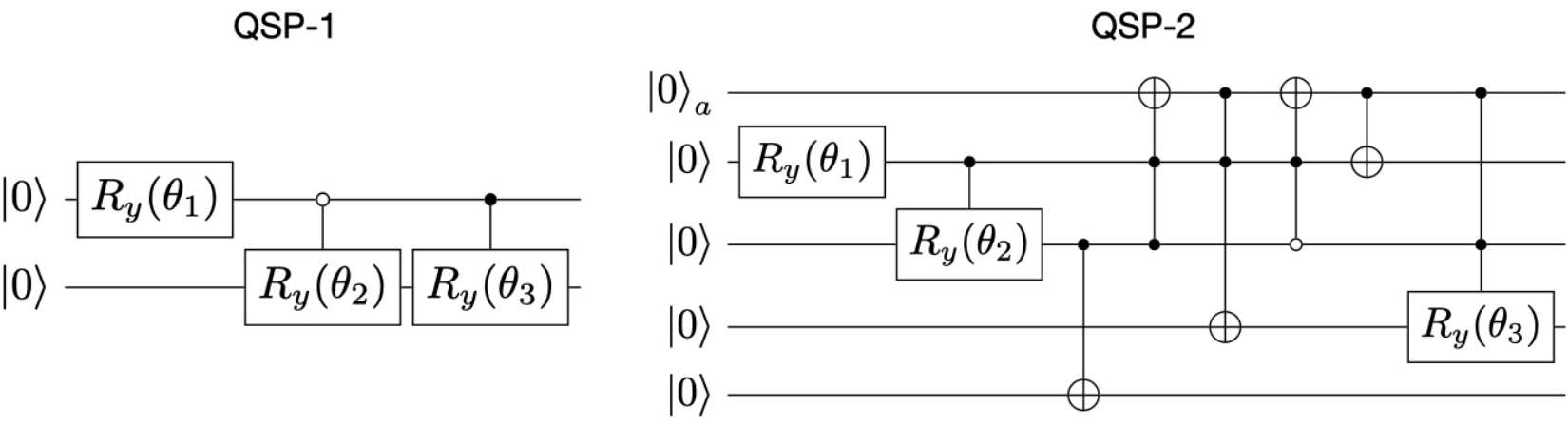}
        }\caption{\justifying (a) Computing the Decision Trees for QSP -1,2 (b) Quantum circuits for QSP -1 and -2 corresponding (a) to load four real values} 
        \label{fig:QSP}
\end{figure*}

Now repeat this $\mathcal{O}(n)$ times to generate an $n$-qubit state with the distribution sampled over $2^{n}$ regions. Though the complexity of such a method could be seen as $\mathcal{O}(n)$, the quantum arithmetic circuits are generally expensive and hence one could alternatively follow an improvement as proposed in \cite{sanders2019black}. To construct a circuit based on the above method described here, we employ the binary tree formulation shown in \cite{prakash2014quantum}, which we shall call QSP-1. Let us consider loading 4 arbitrary non-zero values onto a 2 qubit state as
\begin{equation}
    |\psi\rangle = u_{0}|00\rangle + u_{1}|01\rangle + u_{2}|10\rangle + u_3|11\rangle.
\end{equation}
To construct the circuit with QSP-1, we create a binary tree figure \ref{fig:QSP}(a), left panel, where we start with the values to be loaded as the terminal nodes at the base of the tree. These values are pairwise squared (since probabilities are squares of amplitudes) and summed to build the tree upwards. Finally, all nodes should converge to 1, as they should. Now, to prepare the quantum states, we traverse the tree downwards starting from the vertex, which corresponds to the initial state $|\psi\rangle = |0\rangle\otimes |0\rangle$. Every node has two children: the zero-child and the one-child which have amplitudes corresponding to either the $|0\rangle$ or $|1\rangle$ state, respectively. Then we compute at every node the angle $\theta = \arccos(\sqrt{\frac{\text{zero-child}}{\text{one-child}}})$ and apply a controlled $R_{y}(\theta)$ gate, where the control sequence is the bit sequence of the corresponding node. In the example shown in figure S4, to compute $\theta_{2}$, we look at the two children and obtain 0.25(zero-child) and 0.5(one-child). Therefore $\theta_{2} = \arccos(\sqrt{\frac{0.25}{0.5}})= \pi /4$. Here the parent node, 0.75 corresponds to a zero-child branch. So the controlled-$R_{y}$ gate will operate when qubit 1 is set to 0. Thus, after $R_{y}(\theta_{1})$ we get $\sqrt{0.75}|0\rangle+\sqrt{0.25}|1\rangle)\otimes|0\rangle$. Further with successive applications of controlled gates $R_{y}(\theta_{2})$ and $R_{y}(\theta_{3})$ we obtain
\begin{align}
    &|0\rangle\otimes 0\rangle \xrightarrow{R_{y}(\theta_{1})}\sqrt{0.75}|0\rangle+\sqrt{0.25}|1\rangle)\otimes|0\rangle\\
&\sqrt{0.75}|0\rangle+\sqrt{0.25}|1\rangle)\otimes|0\rangle  \xrightarrow{R_{y} (\theta_{2}),R_{y}(\theta_{3})} \nonumber \\ 
         &\sqrt{0.25}|00\rangle + \sqrt{0.5}|01\rangle  + \sqrt{0.125}|10\rangle + \sqrt{0.125}|11\rangle.
\end{align}
         
\begin{figure*}[htpb!]
        \centering
    \includegraphics[trim={0cm 0cm 0cm 0cm},clip=true,width=1\textwidth]{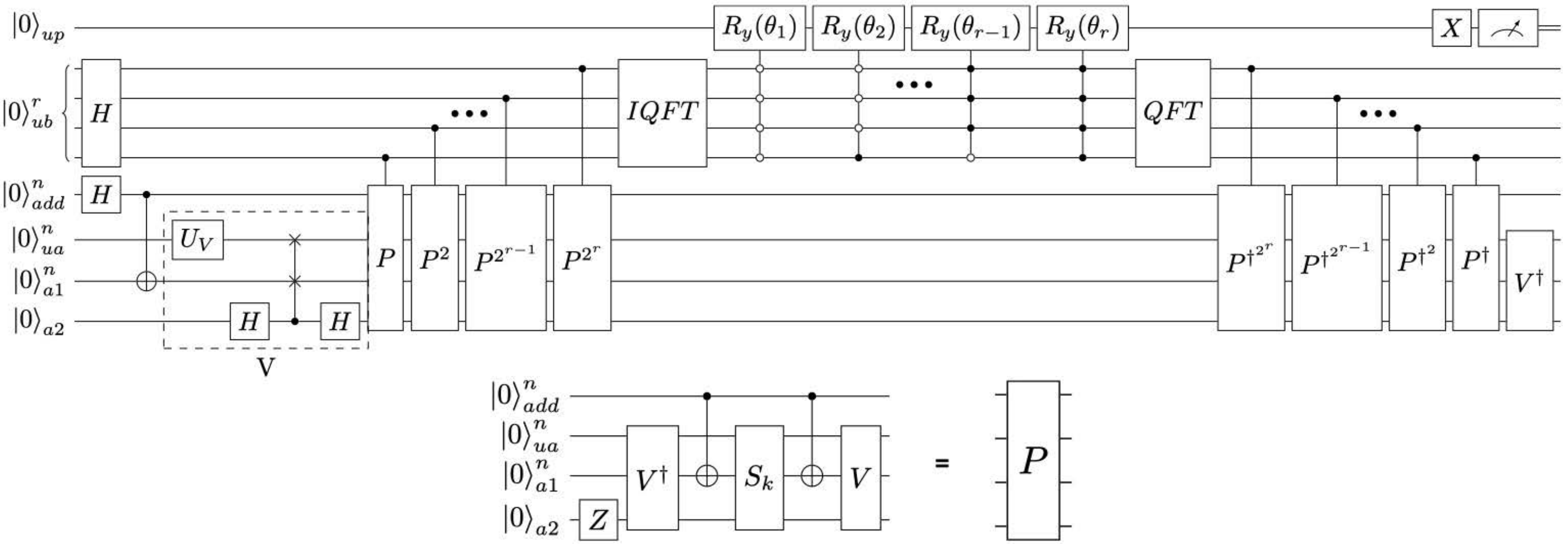}
        \caption{Quantum Post Processing (QPP) circuit}
        \label{fig:QPP circuit}
\end{figure*}
The corresponding quantum circuit for QSP-1 is shown in the left panel of figure \ref{fig:QSP}(b). Now notice that such a method has a total of $\mathcal{O}(n)$ stages which is logarithmic in the size of the state vector. Methods based on \cite{grover2002creating} will work best when the input velocity field to be loaded has a log-concave form. However, when we look at the gate complexity of the number of CNOTs, it grows exponentially as $\mathcal{O}(2^{n})$. To ameliorate this, we employ QSP-2, which drastically reduces the CNOT gate count by $\geq 90\%$; but this is true for sparse quantum states and is based on the method proposed in \cite{mozafari2022efficient,angel}. In fact, this condition works in our favor for the linear solvers considered here. Firstly, we know that between the two time-stepping schemes used in this work, one shot FE and BE schemes are more accurate and efficient since it does not need repeated measurements and state preparation at every time step. The right hand side of the equation in that approach can be readily seen to be a sparse vector; importantly, we consider here all the quantum registers including ancilla qubits to be one single sparse state that needs to be prepared. For flow problems and system sizes considered here, this provides an initial sparse quantum state with $N_{nz}$ non-zero elements, where $ N_{nz}\in [n,5n]$ with an improvement in CNOT count as high as $\approx 94\%$. Further unlike other state preparation algorithms which require exponentially large number of ancillary qubits, this method requires only one. To implement such a circuit, we consider the same 4 values but now loaded onto a 4 qubit state to make it sparse ($N_{nz}=n$). Again we construct a tree called the Decision Diagram (DD) \cite{mozafari2022efficient} as shown in the right panel of figure \ref{fig:QSP}(a). The tree is constructed such that $q_{1}$ through $q_{4}$ depict the different qubits $|q_{1}q_{2}q_{3}q_{4}\rangle$. The solid and dashed arrows/edges denotes whether that specific parent node evaluates 0 or 1, and points to the corresponding zero-child or one-child. For instance, the left-most branch of QSP-2 in figure \ref{fig:QSP}(a) has the sequence of solid-solid-dashed-solid lines, corresponding to the state $|1101\rangle$. However, owing to possible redundant connections one can sometimes invoke reduction rules \cite{mozafari2022efficient} to simplify the DD by eliminating a few nodes; for instance, the right-most node with $q_{3}$ can be simplified as shown, since both children at the terminal nodes point to the same value. Thus by following the rules as detailed in \cite{mozafari2022efficient}, one can generate an efficient circuit with minimal CNOT gates as shown in fig.~\ref{fig:QSP}(b) (right panel), which is of the order $k\times \mathcal{O}(n)<<\mathcal{O}(2^{n})$, where $k$ is the number of paths $\leq N_{nz}$ in the DD, whereas the time complexity is $\mathcal{O}(2kn)$. This procedure drastically reduces the number of CNOT gates thus making it more amenable for implementation on near-term QCs.

\section{Quantum Post Processing}
\label{sec:QPP Append}
In this section we outline the Quantum Post Processing (QPP) protocol for computing the viscous dissipation rate per unit volume $\varepsilon=\nu\langle(\frac{\partial u}{\partial y})^{2}\rangle$; the {angular} brackets indicate a space average. The method proposed here is versatile, making it applicable to more general nonlinear quantities. In brief, to compute the above quantity we would first need to take the derivative of the velocity field with respect to the wall-normal direction $\partial u /\partial y$, for which we employ the well-known spectral method as discussed in Section 4.A. Further, for computing the square of that quantity (or any nonlinear function) we first invoke the Quantum Analog-Digital Converter (QADC-QDAC)\cite{mitarai2019quantum} to convert the representation of the derivatives into binary format, and then to perform either quantum arithmetic or direct controlled rotation operations to finally yield squares of the derivatives in amplitude-encoding format, after undoing the QADC operation outlined in Section 4.B.  

\subsection{Velocity gradients - spectral method}
\label{subsec:spectral gradient}
Computing derivatives in spectral space instead of real space is tantamount to simple scalar multiplication of vector elements by corresponding wave-vectors $k$. {This method, however, applies only for systems with periodic boundary conditions.} Let us consider the following n-qubit state resulting from a QLSA solution 
\begin{equation}
    |\phi\rangle = \sum_{j=0}^{N_{g}-1}u_{j}|j\rangle.
    \label{eq:initial}
\end{equation}
where $u_{i}$ are the velocities at different grid points. First, we apply the Inverse Discrete Fourier Transform (IDFT), which in the quantum setting is the IQFT, to transition into spectral space as 
\begin{equation}
    U_{IQFT}|\phi\rangle \mapsto \frac{1}{\sqrt{N}}\sum_{k=0}^{k=N_{g}}e^{-2\pi i\frac{jk}{N_{g}}}u_{k}|k\rangle.
\end{equation}
Next we multiply the state by a constant diagonal matrix $\Lambda$ defined \cite{an2022efficient} as,
\begin{equation}
\Lambda_{kk} = 
\left\{
    \begin{array}{lr}
        2\pi ik, &  k \in [0,N_{g}/2 -1]\\
        0, & k = N_{g}/2\\
        2\pi i (k-N_{g}), &  k \in [N_{g}/2 +1,N_{g}-1]\\
    \end{array}
\right\} 
\end{equation} and perform the QFT (given by the kernel $e^{2\pi i\frac{jk}{N_{g}}}$) to transform back, which yields a final state that gives the derivative in real space ($u'=\partial u/\partial y$) as
\begin{equation}
    |\phi'\rangle = U_{IQFT}\Lambda U_{QFT}|\phi\rangle = \sum_{j=0}^{N_{g}-1}\frac{du_{j}}{dy}|j\rangle \equiv \sum_{j=0}^{N_{g}-1}u_{j}'|j\rangle
\end{equation}
{ \subsection{Velocity gradients - LCU method}}
\label{subsec:LCU gradient}
{To be able to compute derivatives for systems with more general boundary conditions (for instance, Dirichlet boundary conditions considered in this paper), we now outline a method that is based on a direct computation of finite difference matrices of derivatives. 
Consider the same state as in eq.~(\ref{eq:initial}). Now let $\bar{D}$ be the matrix operator corresponding to the finite difference approximation of the $\partial / \partial y$. The goal now is to implement the following transformation,
\begin{equation}
      \bar{D}|\phi\rangle = \sum_{j=0}^{N_{g}-1}u'_{j}|j\rangle,
\end{equation}
as a quantum circuit, which implies finding a set of unitary operations/gates that can effect the transform due to $\bar{D}$. This can be done by the method of linear combination of unitaries (LCU) \cite{childs2012}, using which one expresses a given matrix $\bar{D}$ as a linear combination of unitary matrices with appropriate coefficients as,
\begin{equation}
    \bar{D} = \sum_{i=0}^{L}\beta_{i}U_{i},
\end{equation}; this weighted sum is used to construct a probabilistic implementation of $\bar{D}$. If such a unitary basis can be found, then the general procedure to implement it as a quantum circuit is outlined as follows:
\begin{enumerate}
    \item Begin with the two registers $\vert\psi\rangle = \vert0\rangle^{\otimes l}\vert\phi\rangle$, where $l=\log_{2}L$. Using an oracle $V$ (that can be constructed following rules of QSP-1,2), prepare the first register in a state proportional to 
    \begin{equation}
    (V\otimes\mathbb{I})\vert0\rangle^{\otimes l}\vert\phi\rangle = \frac{1}{\sqrt{\sum_{i}\beta}}\sum_{i=0}^{L}\beta_{i}\vert i\rangle\vert\phi\rangle
    \end{equation}
    \item Using this first register as control qubits, apply a uniformly controlled operation $G = \sum_{i}|i\rangle \langle i|\otimes U_{i}$ on $\vert\phi\rangle$. Following which, undo the operation of $V$ to reset the first register giving \begin{equation}
        \frac{1}{\sqrt{\sum_{i}\beta}}(V^{\dag}\otimes \mathbb{I})\otimes G\sum_{i=0}^{L}\beta_{i}\vert i\rangle\vert\phi\rangle.
    \end{equation}
    \item Finally, measuring the first register in the appropriate basis yields a state proportional to
    \begin{equation}
        \vert\psi\rangle = |0\rangle^{l}\bar{D}|\phi\rangle_{u} + |\phi\rangle_{\perp}.
    \end{equation} This represents a state in which the desired output exists as a superposition state, thus has a finite probability of being measured (which is generally small). This therefore constitutes a probabilistic implementation of $\bar{D}$. However, by using certain quantum amplitude amplification techniques these probabilities can be improved significantly.
\end{enumerate}
}
\subsection{Nonlinear transform}
\label{subsec:nonlinear transform}
The final step in computing the dissipation involves computing the squares of the velocity gradients, which requires a nonlinear transformation of the quantum amplitudes. Consider a velocity gradient state vector prepared by an oracle $U_{V}$ as shown in figure 1.a in the main text, which either results from a direct quantum state preparation algorithm or the QLSA itself. Our aim is to implement the mapping
\begin{equation}
    |\zeta\rangle = \sum_{k=0}^{N_{g}-1}u'_{k}|k\rangle \mapsto \sum_{k=0}^{N_{g}-1}(u'_{k})^{2}|k\rangle.
    \label{eq:QPP transformation}
\end{equation}
This requires a total of 6 quantum registers as shown in figure \ref{fig:QPP circuit}, where $q_{up}$ is an ancillary qubit to store the amplitude encoding of $(u')^{2}$, $q_{ub}$ stores the $r$-bit basis encoding of $u'$, $q_{add}$ is the address register, $q_{ua}$ encodes the input amplitude encoding of $u$ and $q_{a1}$, $q_{a2}$ are ancillary qubits. Here, we implement a modified version of the Quantum Analog Digital Converter described in \cite{mitarai2019quantum}. 
The steps are as follows:
\begin{enumerate}
    \item STEP 1: Generate the basis superposition for the address qubits by applying Hadamard gates on $q_{add}$ register to yield $\frac{1}{\sqrt{N}}\sum_{s=0}^{N_{add}-1}|s\rangle_{add}$. Following this, we apply CNOT gate on $q_{a1}$ to clone the basis of the address register.
    \item STEP 2: We then load the velocity derivative values into $q_{ua}$ by the oracle $U_{V}$, which gives $\frac{1}{\sqrt{N}}\sum_{s}\sum_{c}u'_{c}|s\rangle_{add}|c\rangle_{ua}|s\rangle_{a1}.$
    \item STEP 3: Since we are interested finally in a squared quantity we can extract simply the absolute value of $u'$ by using the SWAP test circuit (denoted by the portion of circuit enclosed in the dotted box $V$ in figure \ref{fig:QPP circuit}, excluding $U_{V}$). This is a procedure for comparing two states to determine their closeness by estimating the inner product. The SWAP test of two state $|\phi_{1}\rangle$ and $|\phi_{2}\rangle$ yields a quantity of the type $\frac{1}{2}(1-|\langle\phi_{1}|\phi_{2}\rangle|)$. Here we use the test without any measurements, which gives us absolute values of $u'$. This leaves us with
    \begin{align}
        = \sum_{s} \frac{1}{2\sqrt{N}}&|k\rangle_{add} \Big[ \underbrace{\Big(\sum_{c}u'_{c}|c\rangle_{ua}|s\rangle_{a1} +|s\rangle_{ua}\sum_{c}u'_{c}|c_{a1}\rangle\Big)}_{\scalebox{1.2}{$~~~~~~~~~\zeta_{s0}$}}|0\rangle_{a2} \nonumber \\  &+\overbrace{\Big(\sum_{c}u'_{c}|c\rangle_{ua}|s\rangle_{a1} -|s\rangle_{ua}\sum_{c}u'_{c}|c_{a1}\rangle\Big)}^{\scalebox{1.2}{$~~~~~~~~~\zeta_{s1}$}} |1\rangle_{a2}\Big]
        \end{align}
        
\begin{equation}
        = \sum_{s} \frac{1}{\sqrt{N}}|k\rangle_{add} \Big[ \alpha_{s0}|\zeta_{s0}\rangle|0\rangle_{a2}  + \alpha_{s1}|\zeta_{s1}\rangle|1\rangle_{a2}\Big]
\end{equation}

    \item STEP 4: We now perform Quantum Phase Estimation with a gate P defined as shown in fig.~\ref{fig:QPP circuit} (bottom panel), where $S_{k} = \mathcal{I} - 2(|0\rangle\langle 0|)_{ua,a2}\otimes (|s\rangle\langle s|)_{a1}$, this being a conditional phase shift operator. The details of this step can be found in \cite{mozafari2022efficient}. The QPE along with the $IQFT$ results in the state
    \begin{align}
    &|\zeta\rangle_{ub,ua,a1,a2} = \nonumber \\ &\frac{1}{\sqrt{2N}}\sum_{s}|s\rangle_{add}\big(|\beta\rangle_{ub}|\zeta_{+}\rangle) +|(1-\beta)\rangle_{ub}|\zeta_{-}\rangle\big),
    \label{eq:QPP QPE}
    \end{align}
where $\sin(\pi\beta) = \sqrt{0.5(1+(u'_{c})^{2})} $ is stored as an $r$-bit basis representation, and $\zeta_{\pm} = \frac{1}{\sqrt{2}}(|\zeta_{s0}\rangle \pm i|\zeta_{s1}\rangle)$ form the eigen-basis of $P$.

    \item STEP 5: Now, instead of performing quantum arithmetic as shown in \cite{mitarai2019quantum}, we can directly compute the squares of $u'_{c}$ and transform back to amplitude encoding together by applying conditional rotation operators. We introduce another ancillary qubit $q_{up}$ and apply $R_{y}(\theta_{r})$ operators on it (for a given $c=r$), conditioned on the $q_{up}$ qubits, where $\theta = 2\gamma$. Since $u'_{r} = \sqrt{2\sin^{2}\pi \gamma_{r}-1}$, therefore $\sin(2\gamma_{r})=(u'_{r})^{2}$. Thus $R_{y}(\theta_{r})$ performs the operation, $|0\rangle_{up} \mapsto (\sqrt{(1-(u'_{r})^{4})}|0\rangle + (u'_{r})^{2}|1\rangle)$. For a given $r-$basis, these values of $\theta$ can be hard encoded into the circuit. Following this step, we undo the operations on $q_{ua}$, $q_{a1}$ and $q_{a2}$ to set them to $|0\rangle$, which yields the final state
    \begin{equation}
        R\sum_{c=0}^{N}\big(\sqrt{(1-(u'_{c})^{4})})|0\rangle_{up}+ (u'_{c})^{2}|1\rangle_{up}\big)|c\rangle_{add}|0\rangle_{ub,ua,a1,a2}.
    \end{equation}
   When measured in the computational basis after applying $X$ gate, this gives
    \begin{equation}
        R'\sum_{c=0}^{N}(u'_{c})^{2}|0\rangle_{up}|c\rangle_{add}|0\rangle_{ub,ua,a1,a2},
    \end{equation}
    where $R$ and $R'$ are corresponding normalization constants. The above equation is the transformation we sought in eq. \ref{eq:QPP transformation}. Given an n-qubit state such as $|\psi\rangle^{n} = \sum_{p}w_{p}|p\rangle$, we can compute the sum of all amplitudes by applying $U_{avg}=H^{\otimes n}$ which gives $\title{w}_{0} = \sum_{p}w_{p}$. Using this we can compute the sum of, $\sum_{c}(u'_{c})^2$ and finally measure the first qubit basis state with this value. Of course, we would need to post-multiply it by corresponding normalization constants to retrieve the right solution; importantly, we have to divide the final solution by 2, since we know that $\sin (2\pi\gamma) = \sin(2\pi(1-\gamma)$. From eq.~\ref{eq:QPP QPE} we observe that we will get repeated values when we transform into amplitude encoding. Further, we multiply classically with the appropriate viscosity $\nu$ and divide the solution by the number of grid points to yield the final dissipation rate,
    \begin{equation}
       \varepsilon=\nu\Big\langle\Big(\frac{\partial u}{\partial y}\Big)^{2}\Big\rangle.
    \end{equation}
The complexity of the above QPP is as follows: (a) STEPS 1-3 require $\mathcal{O}(\log_{2}N)$ gates; (b) STEP 4 has single and two qubit gates with a complexity of $\mathcal{O}((\log_{2}N)^{2}/\epsilon_{QPP})$ along with $\mathcal{O}(1/\epsilon_{QPP})$ calls to $U_{V}$; (c) STEP 5, which uses controlled rotations has the complexity $\mathcal{O}(1/\epsilon_{QPP})$. Thus, the overall complexity of $\mathcal{O}((\log_{2}N)^{2}/\epsilon_{QPP})$. $U_{V}$ is either the QLSA itself or, if the form of the velocity field is known, it can be prepared by QSP-1,2 thus amounting to a complexity of $ \mathcal{O}(U_{V}) = \min\{\mathcal{O}(\text{poly~log} (N/\epsilon)\kappa /\epsilon_{QPP}), \mathcal{O}(n),\mathcal{O}(kn)\}$. The complexity of the entire algorithm presented in this work is summarized in Table \ref{table:Complexity summary} (where $n=log_{2}(N))$. We caution the readers that these complexities are only estimates and warrant a more detailed analysis of the space, time and gate complexity of the algorithms presented here.
\end{enumerate}
 \begin{table}[htb!]\centering
 \caption{Summary of time complexity of quantum subroutines}

 \begin{tabular}{l|r}
 Quantum subroutine & Complexity \\
 \midrule
 QSP-1 & $\mathcal{O}(n)$  \\
 QSP-2 & $\mathcal{O}(2kn)$ \\
 QLSA-1 & $\mathcal{O}(\text{poly~log}(N)s^{2}\kappa^{2} /\epsilon)$  \\
 QLSA-2 & $\mathcal{O}(\text{poly~log}(N/\epsilon)\kappa )$ \\
 QPP &   $\mathcal{O}(U_{V})+\mathcal{O}((\log N)^{2}/\epsilon_{QPP})$\\
 \midrule
 End-to-end & $\min\{ \mathcal{O}(\text{poly~log} (N/\epsilon)\kappa /\epsilon_{QPP}), \mathcal{O}(kn/\epsilon_{QPP})\}$ \\
  &  $+ \mathcal{O}((\log N)^{2}/\epsilon_{QPP}))$\\
  & $\approx \mathcal{O}(\text{poly~log} (N/\epsilon)\kappa/\epsilon_{QPP})$\\
 \bottomrule
 \end{tabular}
 \label{table:Complexity summary}
 \end{table}
\section{Glossary of acronyms}
\label{sec:glossary}
\noindent 1. ADC - Analog-Digital Converter\\
2. BE - Backward Euler\\
3. CFD - Computational Fluid Dynamics\\
4. DAC -  Digital-Analog Converter\\
5. DD - Decision Diagrams\\
6. DNS -  Direct Numerical Simulations\\
7. FE - Forward Euler\\
8. HHL -  Harrow-Hassidim-Lloyd\\
9. IQFT - Inverse Quantum Fourier Transform\\
10. LCU -  Linear Combination of Unitaries\\
11. NISQ - Noisy Intermediate Scale Quantum\\
12. QC - Quantum Computing\\
13. QCFD - Quantum Computation of Fluid Dynamics\\
14. QFlowS - Quantum Flow Simulator\\
15. QFT - Quantum Fourier Transform\\
16. QLSA - Quantum Linear Systems Algorithms\\
17. QPE - Quantum Phase Estimation\\
18. QPP - Quantum Post Processing\\
19. QSP - Quantum State Preparation

\bibliography{refs}
\end{document}